\documentclass[aps, prb, twocolumn, showpacs, amsmath, amssymb,superscriptaddress]
{revtex4-1}

\usepackage{amssymb,amsfonts,amsmath} 
\usepackage{graphicx}
\usepackage{color}
\usepackage{hyperref}
\usepackage{longtable}
\usepackage{bbm}
\usepackage{nicefrac}

\usepackage{tikz}
\usetikzlibrary{calc}

\newcommand{\PBC}[1]{%
\begin{tikzpicture}[#1]%
\draw[thick] (3mm,0.75mm) circle [x radius=6mm, y radius=1.5mm];
\end{tikzpicture}%
}

\newcommand{\PBCbvthree}[2]{%
\begin{tikzpicture}[#1]%
\draw[thick] (3mm,0.75mm) circle [x radius=6mm, y radius=1.5mm];
\filldraw[fill=white, draw=white] (-1mm,1mm) rectangle (7mm,3mm) ;
\draw[draw=none] (-1mm,-0.25mm) rectangle (3mm,2.25mm) node[align=center] {\tiny {\color{black}$b\!\!=\!\!#2$}};

\end{tikzpicture}%
}

\newcommand{\PBCblone}[2]{%
\begin{tikzpicture}[#1]%
\draw[thick] (3mm,0.75mm) circle [x radius=6mm, y radius=1.5mm];
\filldraw[fill=white, draw=white] (1mm,1mm) rectangle (5mm,3mm) ;
\draw[draw=none] (-1mm,-0.25mm) rectangle (3mm,2.25mm) node[align=center] {\tiny {\color{black}$b\!\!>\!\!#2$}};

\end{tikzpicture}%
}

\newcommand{\PBCbvone}[2]{%
\begin{tikzpicture}[#1]%
\draw[thick] (3mm,0.75mm) circle [x radius=6mm, y radius=1.5mm];
\filldraw[fill=white, draw=white] (1mm,1mm) rectangle (5mm,3mm) ;
\draw[draw=none] (-1mm,-0.25mm) rectangle (3mm,2.25mm) node[align=center] {\tiny {\color{black}$b\!\!=\!\!#2$}};

\end{tikzpicture}%
}

\newcommand{\PBCb}[1]{%
\begin{tikzpicture}[#1]%
\draw[thick] (3mm,0.75mm) circle [x radius=6mm, y radius=1.5mm];
\filldraw[fill=white, draw=white] (2mm,1mm) rectangle (4mm,3mm) ;
\draw[draw=none] (-1mm,-0.5mm) rectangle (3mm,2.5mm) node[align=center] {\tiny {\color{black}$b$}};
\end{tikzpicture}%
}

\newcommand{\OBC}[1]{%
\begin{tikzpicture}[#1]%
\draw[thick] (0mm,0.75mm) -- (10mm,0.75mm);
\end{tikzpicture}%
}

\newcommand{\OBCb}[1]{%
\begin{tikzpicture}[#1]%
\draw[thick] (0mm,0.75mm) -- (10mm,0.75mm);
\filldraw[fill=white, draw=white] (4mm,-0.mm) rectangle (6mm,2mm) ;
\draw[draw=none,anchor=north east] (2.75mm,-0.75mm) rectangle (6.75mm,2.75mm) node[align=center] {\tiny {\color{black}$b$}};
\end{tikzpicture}%
}

\newcommand{\OBCbvone}[2]{%
\begin{tikzpicture}[#1]%
\draw[thick] (0mm,0.75mm) -- (12mm,0.75mm);
\filldraw[fill=white, draw=white] (4mm,-0.mm) rectangle (8.5mm,2mm) ;
\draw[draw=none,anchor=north east] (2.75mm,-0.75mm) rectangle (9.25mm,2.75mm) node[align=center] {\tiny {\color{black}$b\!\!=\!\!#2$}};

\end{tikzpicture}%
}

\newcommand{\OBCblone}[2]{%
\begin{tikzpicture}[#1]%
\draw[thick] (0mm,0.75mm) -- (12mm,0.75mm);
\filldraw[fill=white, draw=white] (4mm,-0.mm) rectangle (8.5mm,2mm) ;
\draw[draw=none,anchor=north east] (2.75mm,-0.75mm) rectangle (9.25mm,2.75mm) node[align=center] {\tiny {\color{black}$b\!\!>\!\!#2$}};

\end{tikzpicture}%
}

\newcommand{\ve}{\mathbf}

\begin{document} \title{Renormalization group flows in one-dimensional lattice models: \\ 
impurity scaling, umklapp scattering and the orthogonality catastrophe}

\author{D.M.\ Kennes}  
\affiliation{Institut f{\"u}r Theorie der Statistischen Physik, RWTH Aachen University 
and JARA---Fundamentals of Future Information
Technology, 52056 Aachen, Germany}

\author{M.J.\ Schmidt}  
\affiliation{Institut f{\"u}r Theoretische Festk\"orperphysik, RWTH Aachen University, 52056 Aachen, 
Germany}

\author{D. H\"ubscher}  
\affiliation{Institut f{\"u}r Theoretische Festk\"orperphysik, RWTH Aachen University, 52056 Aachen, 
Germany}

\author{V.\ Meden} 
\affiliation{Institut f{\"u}r Theorie der Statistischen Physik, RWTH Aachen University 
and JARA---Fundamentals of Future Information
Technology, 52056 Aachen, Germany}

\begin{abstract} 

We show that to understand the orthogonality catastrophe in the half-filled
lattice model of spinless fermions with repulsive nearest neighbor
interaction and
a local impurity in its Luttinger liquid phase one has to take into
account (i)
the impurity scaling, (ii) unusual finite size $L$ corrections of the
form $\ln(L)/L$,
as well as (iii) the renormalization group flow of the umklapp
scattering. The
latter defines a length scale $L_u$  which becomes exceedingly large the
closer the system
is to its transition into the charge-density wave phase. Beyond this transition umklapp
scattering is
relevant in the renormalization group sense. Field theory can only be
employed for length scales
larger than $L_u$.  For small to intermediate two-particle interactions,
for which the regime
$L > L_u$ can be accessed, and taking into account the finite size
corrections resulting from (i) and (ii)
we provide strong evidence that the impurity backscattering
contribution to the orthogonality exponent is asymptotically given by
$1/16$. While further increasing the two-particle interaction leads to a
faster
renormalization group flow of the impurity towards the cut chain fixed
point, the increased bare amplitude of the umklapp scattering renders it
virtually
impossible to confirm the expected asymptotic value of $1/16$ given the
accessible
system sizes. We employ the density matrix renormalization group.

\end{abstract}

\date{\today} 

\maketitle

\section{Introduction}

Early indications that a single local impurity has dramatic effects on the low-energy physics 
of a one-dimensional (1D) Luttinger liquid (LL)\cite{Luther74,Mattis74} were phrased in 
the modern language of renormalization group (RG) relevance and RG flows in the seminal work 
of Ref.~\onlinecite{Kane92}. Considering the field theoretical Tomonaga-Luttinger model
(TLM)\cite{Giamarchi03,Schoenhammer05} and using perturbative RG in the impurity strength as 
well as the 
amplitude of a weak hopping between two open chains it was shown that for repulsive 
two-particle interactions a weak impurity with a 
{\em finite backscattering contribution} is a relevant perturbation, 
while a weak hopping is RG irrelevant.\cite{Kane92}  The RG flow from the perfect to the 
cut chain fixed points within the continuum TLM was later confirmed by nonperturbative 
approaches.\cite{Moon93,Matveev93,Fendley95} These works were mainly concerned with 
transport and spectral properties of inhomogeneous LLs but, 
soon after,
other quantities indicative of the impurity RG flow were investigated as well. 

A rather fundamental one is the overlap $O$ between the ground state of the homogeneous 
 system and the one of the same system supplemented by a single local impurity. As 
shown 
by Anderson,\cite{Anderson67} for noninteracting fermions this overlap vanishes as a function of 
the 
system size $L$ following a power law $O\sim L^{-\alpha}$. The {\em orthogonality exponent} (OE) 
$\alpha >0$ of this {\em orthogonality catastrophe} is fixed by the scattering phase shifts of 
the impurity.\cite{Mahan90} It enters the exponents of edge singularities in x-ray spectra 
of metals\cite{Mahan90} as well as the low-energy properties of prototypical quantum dot models
such as the interacting resonant level model in and out of 
equilibrium.\cite{Schlottmann80,Kashcheyevs09,Muender12,Doyon07,Karrasch10}

If the potential of the bare impurity 
varies weakly on the scale of $k_F^{-1}$, 
such that the 
backscattering vanishes, the changes of the OE due to {\em two-particle interactions} can be 
computed 
exactly\cite{Ogawa92,Lee92} within the 1D continuum TLM using 
bosonization.\cite{Schotte69,Giamarchi03,Schoenhammer05} Here $k_F$ denotes the Fermi wave 
vector. 
The amplitude of the forward impurity scattering
does not flow and thus the forward scattering contribution to the OE is only weakly 
affected by interactions.\cite{Ogawa92,Lee92} Furthermore, due to the linearization of the 
single-particle 
dispersion inherent to the construction of the TLM\cite{Giamarchi03,Schoenhammer05} this contribution reduces to the Born approximation for 
the 
forward scattering phase shift at vanishing two-particle interaction. Bosonization thus only makes 
a prediction for the forward scattering contribution of the OE for very weak impurities.
\cite{Meden98} 
In the present work we do not consider impurity forward scattering. Instead, we study a particle-hole 
symmetric lattice model which is tailored such that forward scattering vanishes.\cite{Meden98}

In the presence of even a small impurity backscattering, however, the consequences of the impurity 
RG flow towards the cut chain fixed point are striking: on low energy scales, that is for 
large system sizes $L$, even a weak impurity effectively acts as an open boundary leading to a 
value 
$\alpha=1/16$ of the 
OE---related to a phase shift of $\pm \pi/2$---which is independent of the bare impurity strength and the two-particle 
interaction of the TLM.\cite{Gogolin93,Prokofev94,Kane94,Affleck94,Komnik97,Furusaki97}  

Early attempts to confirm the impurity RG scaling close to the perfect and cut chain fixed 
points in {\em microscopic lattice models} using exact diagonalization\cite{Eggert92} were latter 
complemented by functional RG\cite{Metzner12} results which reveal the full crossover flow 
for spectral and transport properties.\cite{Andergassen04,Enss05} The expectation that the 
impurity RG flow 
of the TLM supplemented by a local impurity should also be observable in lattice models is based 
on the observation that the {\em translational invariant} TLM forms the low-energy fixed point 
model of a large class of 
homogeneous 1D metallic Fermi systems. This lies at the heart of LL universality.\cite{Haldane80} 
To show this type of universal behavior one has to understand the RG flow of different 
{\em two-particle scattering processes} (g-ology model).\cite{Solyom79} In lattice 
models of spinless fermions, which we consider here, particular 
attention has to be paid to umklapp scattering. More specifically, we study the lattice model of 
spinless 
fermions with nearest neighbor hopping $t$ and  nearest neighbor interaction $U$ at half filling. 
This model falls into the LL universality class for $-2 < U/t < 2$. Within this parameter regime 
umklapp scattering is RG irrelevant. It becomes relevant for $U/t>2$ leading to the transition 
into a charge-density wave  state.\cite{Giamarchi03,Schoenhammer05}  

For lattice models with repulsive two-particle interaction the value $1/16$ for 
$\alpha$ was so far not convincingly demonstrated. Early 
attempts using density matrix renormalization group (DMRG) pointed towards this 
value.\cite{Qin96,Qin97} However, closer inspection within a comprehensive DMRG 
study using larger systems with up to 100 lattice sites showed that 
the results were inconclusive; see Fig.~5 of Ref.~\onlinecite{Meden98}. 
We here revisit this problem. Our study is based on  (a) recent field theoretical insights on the 
overlap of a 1D system,\cite{Dubail11,Stephan11,Stephan13} 
(b) the finding that umklapp scattering cannot be ignored for $U/t \gtrapprox 1$, as well as (c) 
the progress in computer speed and the DMRG algorithm.\cite{Schollwoeck11} 

DMRG combines two advantages vital for our investigation: 
it provides highly accurate ground state wave functions 
and allows to study larger systems than obtainable by any other `numerically  exact' approach 
to 1D quantum many-body systems.

In a series of papers\cite{Dubail11,Stephan11,Stephan13} it 
was argued that the logarithm of 
the ground state overlap of an open chain and an open chain additionally cut in the 
middle (infinite impurity strength) can be viewed as the {\em free energy} of a 
1+1-dimensional classical boundary problem as long as the chain can be described 
by a field theory. Based on this it was shown that this specific overlap is 
characterized by unusual finite size corrections of the form  $\ln(L)/L$. We here 
show that these corrections are also crucial to understand the system 
size dependence of the overlap in 
our interacting lattice model including the case of a {\em finite} impurity and thus the OE in general. We first study the overlap of the ground states of 
an open chain and an open chain with a bond impurity in the middle and second the 
one of a periodic chain and a periodic chain with a bond impurity. Considering 
the logarithm of the overlap as a free energy provides the justification to add a 
typical impurity scaling term\cite{Kane92} $\sim L^{1-1/K}$ to the finite size scaling 
of $\ln |O|$,\cite{thanks} with $1/K$ being the scaling dimension of the residual hopping 
close to the cut chain fixed point and $K \leq 1$ the LL parameter of the lattice 
model which depends on $U/t$.\cite{Giamarchi03,Schoenhammer05} 
By taking these finite size corrections as well as a standard term $\sim 1/L$ 
into account our results of the OE  turn out to be consistent with 
the asymptotic $\alpha=1/16$ for $U/t \lessapprox 1$. Interestingly, the interplay of the impurity 
scaling $\sim L^{1-1/K}$  and the $\ln(L)/L$ correction leads to highly unusual finite size scaling of 
the backscattering component of the OE.
It should however be noted that, although we provide evidence that the asymptotic OE is 1/16 for $U>0$, the system sizes corresponding to the asymptotic regime cannot be reached for small $U/t$, neither in numerical simulations nor in actual experiments.

As $K$ decreases with increasing interaction strength $U$
one is tempted to consider two-particle interactions close to the transition into 
the charge-density wave phase at $U/t=2$ for which $K=1/2$. In this 
limit the finite size corrections by the impurity flow vanish faster. However, for $U/t \to 2$ the 
amplitude 
of the flowing umklapp scattering at the largest accessible system sizes (up to a few thousand lattice sites) 
is still too large to be negligible and field theory cannot be employed. This renders it virtually 
impossible to 
conclusively 
demonstrate the asymptotic value $\alpha=1/16$ for $U/t\gtrapprox 1$. 
The umklapp scattering defines a length scale $L_u$ which strongly increases the closer 
the system comes to its phase transition. In short, to understand the orthogonality catastrophe 
in our lattice model for the accessible system sizes of up to a few thousand lattice sites one 
has to consider both the {\em single-particle} impurity RG flow\cite{Kane92} as well as the 
flow of components of the {\em two-particle} interaction.\cite{Solyom79} The appearance of 
a scale which restricts field theoretical behavior, e.g. typical LL power laws, to exceedingly
large systems the larger $U$ was earlier shown---but not fully analyzed---for the momentum $k$ distribution 
function $n(k)$ of our translational invariant lattice model.\cite{Karrasch12} To complement our 
results for $L_u$ extracted from the overlap we repeat this study and provide evidence that 
also this scale stems from umklapp scattering.

The remainder of this paper is organized as follows. In Sect.~\ref{sect_model} we introduce our 
lattice model and briefly discuss the weak coupling RG flow of umklapp scattering 
(g-ology)\cite{Solyom79,Giamarchi03} as well as prior results on the impurity  scaling 
obtained for the microscopic model. We discuss basics on wave function overlaps, their finite
size dependence, and our way of analyzing the numerical data for $O$ in Sect.~\ref{sect_over}. 
In Sect.~\ref{sect_umklapp_scale} we relate the umklapp scales $L_u$ extracted from the 
$L$-dependence of the overlap, the $k-k_F$-dependence of the momentum 
distribution of the translational invariant lattice model as well as the weak coupling RG of the continuum 
g-ology model. Section \ref{sect_weak_link} contains our DMRG results of the OE for systems with 
open (OBC) and periodic boundary conditions (PBC). Our results are summarized in Sect.~\ref{sect_sum}. 
The appendices \ref{ap_variable} and \ref{ap_DMRG} contain details of our fitting procedures and 
the DMRG implementation, respectively.

\section{Model}

\begin{table}
\begin{ruledtabular}
\begin{tabular}{ccc}
  symbolic representation &  OBC & PBC  \\ \hline
 perfect chain $b=1$       & $\OBC{baseline=height}$           & $\PBC{baseline=height}$  \\
 impurity $b< 1$   & $\OBCb{baseline=height}$ & $\PBCb{baseline=height}$  
\end{tabular}
\end{ruledtabular}
\caption{Symbols used for the different setups featuring OBC and PBC each with $b=1$ (perfect 
chain) or 
$b< 1$ (hopping impurity at center of chain). \label{tab:shorthand}} 
\end{table}
\label{sect_model}

In the following we consider interacting spinless fermions on a lattice 
(lattice constant $a=1$) 
described by the Hamiltonian
\begin{equation}
\begin{split}
H=&-t\sum_j \left[ c^\dagger_{j+1}c^{\phantom \dagger}_{j}+c^\dagger_{j}c^{\phantom \dagger}_{j
+1}\right]\\&+ U \sum_j 
\left[\left(c^\dagger_{j}c^{\phantom \dagger}_{j}-\frac12\right)\left(c^\dagger_{j+1}c^{\phantom 
\dagger}_{j+1}-\frac12\right)\right]
\end{split}
\label{eq:Lattice_Ham}
\end{equation}
in standard second quantization notation. We restrict ourselves to half-filling in this work. The parameters $t>0$ and 
$U (\geq 0)$ determine 
the hopping amplitude between neighboring sites and the density-density type of (repulsive) 
interaction of adjacent 
particles, respectively. We investigate OBC as well as PBC. 
For OBC the sums in 
Eq.~\eqref{eq:Lattice_Ham} run from sites $j=1$ to $j=L-1$, while for PBC the upper bound of the 
sum is given 
by $j=L$ with $c^{(\dagger)}_{L+1}=c^{(\dagger)}_1$.

The above Hamiltonian is supplemented by a bond impurity 
\begin{equation}
\begin{split}
 H_{\rm imp}  & =(1-b)t \left[ c^\dagger_{L/2+1}c^{\phantom \dagger}_{L/2}+c^\dagger_{L/2}
c^{\phantom \dagger}_{L/2+1}\right] 
\\&+ (1-b) U \left[\left(c^\dagger_{L/2}c^{\phantom \dagger}_{L/2}-\frac12\right)\left(c^\dagger_{L/
2+1}c^{\phantom \dagger}_{L/2+1}
-\frac12\right)\right]
\end{split}
\label{eq:Impurity_Ham}
\end{equation}
such that $b=0$ ($b=1$) corresponds to a cut (perfect) chain. For future reference we use 
symbols for 
the four different cases of OBC and PBC each with and without an impurity ($b< 1$) as 
introduced in 
Table \ref{tab:shorthand}. Note that for half band filling the above bond impurity has vanishing 
forward scattering (particle-hole symmetry).\cite{Meden98}

The low-energy physics ($L\to\infty$) of the impurity free model defined in Eq.~
\eqref{eq:Lattice_Ham} is 
known to be characterized by the g-ology model.\cite{Solyom79,Giamarchi03} In this continuum 
model only 
the linear part of the dispersion around the Fermi points as well as the dominant low-energy 
interaction processes 
in compliance with energy and momentum conservation are kept. The linearization of the 
dispersion leads 
to branches of left  ($k\approx-k_{\rm F}$) and right ($k\approx k_{\rm F}$) moving fermions.  After 
performing 
the continuum limit and linearizing the dispersion relation for the model given in  Eq.~
\eqref{eq:Lattice_Ham} 
one can classify different interaction processes. One involves two fermions on the same branch 
denoted as 
$g_4$ and one two particles on different branches denoted as $g_2$. Note that in the present 
spinless 
case $g_2$ processes
with small momentum transfer and $g_1$ processes with momentum transfer $2 k_F$ are 
indistinguishable. The latter 
thus need not be introduced. The $g_4$ and $g_2$ processes conserve momentum. 
Using standard bosonization\cite{Giamarchi03,Schoenhammer05} the Hamiltonian 
containing these two-particle scattering processes is given by a {\em free bosonic field theory},
the TLM, 
\begin{equation}
H_0=\frac{1}{2\pi}\int\;dx\left\{ vK[\partial\theta(x)]^2+\frac{v}{K}[\partial\phi(x)]^2\right\} .
\label{eq:TLM}
\end{equation}
Here $\theta(x)$ and $\phi(x)$ are bosonic fields and the model parameters are the charge velocity 
$v$  and  
the dimensionless LL parameter $K$. 
Using the above described `constructive' bosonization of the lattice model it is only possible to 
extract 
the $U$ and $t$ dependence of $v$ and $K$ for $U/t \ll 1$.

For the lattice model Eq.~\eqref{eq:Lattice_Ham} at half filling one 
additionally encounters an umklapp scattering term $g_3$, for which momentum is 
conserved only up to a vector of the reciprocal lattice. This gives rise to an interacting 
contribution ($y \sim g_3$)
\begin{equation}
H_{u}\sim y\int \;dx \cos[4\phi(x)]
\label{eq:H_um}
\end{equation}
to the bosonized Hamiltonian $H=H_0+H_{u}$ leading to the sine-Gordon 
model.\cite{Giamarchi03} Umklapp scattering breaks scale invariance and the large 
amount of results for the TLM associated to 
the latter are not applicable for $y \neq 0$. Fortunately, using a {\em weak coupling  RG} 
treatment of the umklapp scattering term, with $y \sim U$ assumed to be small, one can 
show, 
that it is RG irrelevant. As the details of the RG flow do not matter in the present section the 
corresponding flow equations\cite{Giamarchi03,Solyom79} are given in Eq.~\eqref{flowequations} below. 
Under the RG flow, 
that 
is for decreasing energy scales (increasing length scales), the umklapp scattering is renormalized 
to 
zero and scale invariance is restored. 
At the end of the flow, where the flow parameter $l  \to \infty$, an 
approximation to the renormalized value of $K$  is obtained by $K(\infty)$. However, it is a priori 
not clear how small the energy scales of a given microscopic model, e.g. our lattice model 
Eq.~\eqref{eq:Lattice_Ham}, has to be, such that the umklapp 
scattering contribution can be safely neglected. Furthermore, the perturbative nature of the RG 
treatment restricts its range of validity to small interaction strength ($U/t \ll 1$). 

Fortunately, for the model of Eq.~\eqref{eq:Lattice_Ham} one can find an exact solution via Bethe 
ansatz. The exact values of $v$ and $K$ at half-filling and $|U|<2t$ can be extracted from 
the Bethe ansatz expression for the ground state energy and read
\cite{Giamarchi03,Schoenhammer05}
\begin{equation}
\begin{split}
K_B&=\frac{\pi}{4\eta} ,\\
v_B&=t\frac{\pi\sin(2\eta)}{\pi-2\eta} ,\\
2\eta&={\rm arccos}\left(-\frac{U}{2t}\right).\label{eq:Kandv_bethe}
\end{split}
\end{equation}
For interactions $0\leq U/t<2$, $K_B$ assumes values $\nicefrac{1}{2}<K_B\leq 1$. By    
a series expansion of $K_B$ in $U/t$ one recovers to leading order the results obtained for $K(\infty)$ 
at the end of the flow of the perturbative RG. Furthermore, $y_B=0$ is found in accordance 
with $y(\infty)=0$ of the weak coupling RG. For $U/t>2$ umklapp scattering turns RG relevant, signaling the 
phase transition to a charge-density wave state. This transition  (at $K=\nicefrac{1}{2}$) is also 
captured by the perturbatively motivated RG equations, although they cannot be used to determine 
$K(\infty)$ and $y(\infty)$ any longer as $y$ flows to strong coupling. 
To summarize, for $U/t\leq 2$ the g-ology RG equations \eqref{flowequations} describe qualitatively 
(and for small $y$ even quantitatively) the fate of the umklapp scattering and its effect on the LL 
parameter 
$K$ of the impurity free lattice model.

A comprehensive picture of the spectral and transport properties of our lattice model 
Eq.~\eqref{eq:Lattice_Ham} supplemented by a 
local impurity, such as e.g. $H_{\rm imp}$ Eq.~\eqref{eq:Impurity_Ham},  at small to intermediate 
interactions 
was obtained using functional RG.\cite{Metzner12,Andergassen04,Enss05} The effect of the 
perfect and cut 
chain fixed points, the corresponding scaling dimensions, as well as the full crossover flow on the 
corresponding observables (local single-particle spectral function, linear conductance) found within the functional-RG approach
are in accordance with the results derived from the TLM Eq.~\eqref{eq:TLM} supplemented by 
an impurity.\cite{Kane92}  The latter corresponds to the {\em local} sine-Gordon Hamiltonian.
\cite{Giamarchi03}
In particular, the effect of a hopping between two decoupled chains vanishes as 
$\Lambda^{1/K-1}$, were $\Lambda$ denotes an infrared cutoff, such as e.g. temperature or 
inverse 
system size $L^{-1}$. In the applied approximation, which is controlled for $U/t \lessapprox 1$, the 
functional 
RG has the distinct advantage that very large systems of up to $10^7$ lattice sites and thus very 
low energy scales are accessible. However, it cannot directly be employed to compute the overlap 
$O$ as it does not aim at wave functions but rather 
$n$-particle Green functions. We note in passing that the approximate functional RG 
method does not capture the phase transition at $U/t=2$ and thus the divergence of the scale $L_u$. 
We here resort to a different approach and use DMRG. For details of our DMRG implementation, 
see Appendix \ref{ap_DMRG}.

\section{Wave function overlaps}
\label{sect_over}

The central quantity to study in the context of the OC is the overlap $O$ of two ground state wave 
functions---one of a system with an impurity $(b< 1)$ and one of a perfect system ($b=1$). In 
this work we study two types of overlaps which differ by the boundary conditions of the 
models. In the case of open boundary conditions, we consider the overlap 
$O=\langle\OBC{baseline=height} | \OBCb{baseline=height}\rangle$ between ground states of a perfect chain and a 
chain with a bond impurity $b< 1$ in the center. For periodic boundary conditions, we study 
the overlap $O=\langle\PBC{baseline=height}|\PBCb{baseline=height}\rangle$. These overlaps depend on the system size $L$ in 
a characteristic way. The OC owes its name to the limiting behavior $\lim_{L\rightarrow\infty} O = 
0$, 
i.e., the fact that the ground states of two infinitely large systems are zero even though they 
differ only by the presence of a local impurity.\cite{Anderson67,Mahan90}

We are particularly interested in the approach of this limit when the systems under consideration 
are still finite. For a noninteracting system one can show\cite{Anderson67,Mahan90} that 
\begin{equation}
O \sim L^{-\alpha} , \;\;\; \alpha>0.
\end{equation}
At $U=0$ the OE $\alpha$ of our model given by $H+H_{\rm imp}$ Eqs.~\eqref{eq:Lattice_Ham} 
and 
\eqref{eq:Impurity_Ham} at half filling can be computed using scattering theory\cite{Meden98}  
\begin{equation}
\alpha_{U=0} = \frac1{4\pi^2} \arcsin^2\left(\frac{1-b^2}{1+b^2}\right).
\label{eq:OE0}
\end{equation}
Combining this result with the expectation that in the interacting model $b$ effectively 
approaches 0 (cut chain fixed point) for large systems one can conjecture that
\begin{equation}
\alpha_{U>0} = \frac{1}{16}.
\end{equation}
However, it remains to be shown that this is indeed the case. Earlier attempts to do so using 
DMRG 
and systems of up to 100 sites indicated a tendency towards this value but turned out to be 
inconclusive under closer inspection.\cite{Qin96,Qin97,Meden98}

In order to thoroughly investigate the finite-size behavior of $O$ and to establish the OE from 
numerical calculations of wave function overlaps in finite size systems, it is customary to study 
the logarithmic derivative of $|O|$. From the DMRG calculations of $O(L)$ we derive the logarithmic derivative
\begin{equation}
D(L+\Delta L/2) = - \frac{\ln |O(L+\Delta L)| - \ln |O(L)|}{\ln(L+\Delta L) - \ln(L)}. 
\label{discrete_log_der}
\end{equation}
For $L\rightarrow\infty$, $D(L)$ should thus converge to the correct OE $\alpha$. It turns out, however, that supporting 
this statement with numerical data is exceedingly difficult without further knowledge about the finite-$L$ functional form of $D(L)$. In the following we dicuss two unusual terms in the finite-size scaling, which are of utmost importance for a conclusive analysis of the numerical data.

\begin{figure}
\centering
\includegraphics[width=\linewidth]{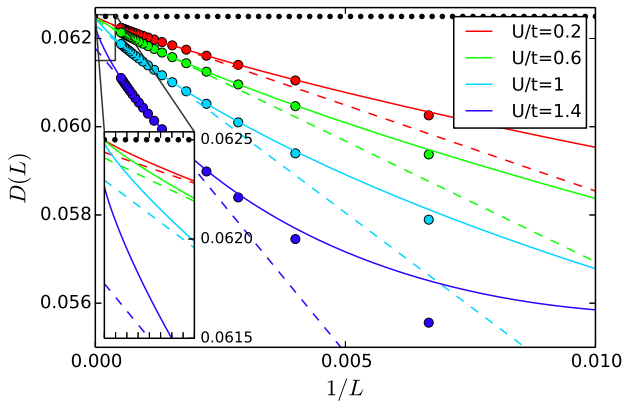}
\caption[]{(Color online) Logarithmic derivatives $D(L)$ of the overlap $O=\langle \OBC{baseline=height} | \OBCbvone{baseline=height}{0}
\rangle$. The dots represent the numerical data for systems with size $L=100,200,\dots,2000$. 
The full lines correspond 
to the fits with the recently-discovered $\ln(L)/L$ term [see Eq. (\ref{fitform})], while the dashed 
lines are the fits without the log term. Only the numerical data for $L>600$ is taken into account for 
the fits. The dotted horizontal line shows the expected $1/16$ limit.}
\label{fig_b0_logform}
\end{figure}

A series of recent conformal field theory (CFT) studies\cite{Dubail11,Stephan11,Stephan13} provided
surprising insights on the finite-$L$ scaling of $\ln |O(L)|$ and thus $D(L)$ for 1D interacting
field theories. More specifically the authors studied the overlap of an open chain and 
an open chain cut in the middle, that is $O=\langle \OBC{baseline=height} | \OBCbvone{baseline=height}{0}
\rangle$ in our 
short hand notation. The impurity scaling does not play a role for this setup as one already starts
at the cut chain fixed point with $b=0$. The authors of Refs.~\onlinecite{Dubail11,Stephan11,Stephan13} 
argued that  $\ln |O|$ can be viewed as the free energy of a 1+1-dimensional classical boundary 
problem. From this they extracted the leading behavior $\ln |O(L)| \sim -\frac{1}{16} \ln(L)$ and 
showed 
that the stress tensor at the boundary leads to the unusual leading finite size correction 
$\ln(L)/L$ which must be supplemented by a regular $1/L$ term. 
For $D(L)$ we thus expect to find 
\begin{equation}
D(L) \approx \alpha + \beta \frac{\ln L}L + \lambda \frac1L 
\label{fitform}
\end{equation}
with $\alpha=1/16$
when studying the same overlap  $\langle \OBC{baseline=height} | \OBCbvone{baseline=height}{0}
\rangle$ in our lattice model. 
Indeed, for $U/t\lessapprox1$ a fit of Eq. (\ref{fitform}) to
the numerical DMRG data for up to 2000 lattice sites extrapolates 
nicely to the correct $\alpha=1/16$  as shown in Fig.~\ref{fig_b0_logform}; 
compare solid 
lines and circles. Note the impressive accuracy of the extrapolation visible in 
the inset (for $U/t\lessapprox1$). The dashed lines in Fig.~\ref{fig_b0_logform} show fits of the DMRG data for $D(L)$ to  
the 
form $\alpha + \lambda/L$. The quality of the fits is clearly worse and in particular does not allow 
to correctly predict the asymptotic value of $1/16$.  The confirmation of the unusual finite size 
corrections predicted by CFT for the logarithm 
of the overlap  $\langle \OBC{baseline=height} | \OBCbvone{baseline=height}{0}
\rangle$ constitutes our first important result.  

However, for $U/t \gtrapprox 1$ even the extrapolations with Eq.~\eqref{fitform} become worse as 
is evident from the lower solid lines of the inset of Fig.~\ref{fig_b0_logform}. An impurity 
of strength $b>0$ can be expected to further increase the relevance of finite size 
corrections in $D(L)$ and we already now conclude that there is not much hope to convincingly 
demonstrate 
$\alpha=1/16$ for $b>0$ and $U/t \gtrapprox 1$ based on data with few thousand lattice sites 
which constitutes
the upper bound reachable with state of the art numerics. 
This constitutes our second important finding.
In the following section we show that this failure originates from sizable umklapp scattering 
and thus the nonapplicability of CFT to the 
strongly interacting regime of the lattice model in too short systems. The minimum system size needed for connecting to CFT results diverges exponentially as the critical point $U/t=2$ is approached.

Viewing $\ln|O|$ as a free energy also provides solid justification to add yet another perturbation to 
the 
finite size scaling of $D(L)$ when studying a {\em finite} bond impurity with $0 < b < 1$. One can 
expect\cite{thanks} that close to the cut chain fixed point 
the impurity contributes with a typical scaling term\cite{Kane92} $\sim L^{1-1/K}$ leading to 
\begin{equation}
D(L) \approx \alpha + \beta \frac{\ln L}L + \lambda \frac1L + \kappa L^{1-1/K} . \label{fitform_kf}
\end{equation}  
In Sect.~\ref{sect_weak_link} we show that this form indeed allows for convincing fits of 
our DMRG data from which $\alpha =1/16$ can eventually be concluded even for $b>0$ and 
small to intermediate interactions $U/t \lessapprox 1$. 

\section{The umklapp scale $L_u$}
\label{sect_umklapp_scale}

The finite size corrections to the OE of the form of Eqs. (\ref{fitform})
and \eqref{fitform_kf} can be applied if the relevant physics is described by 
a CFT. In the bosonic representation of the lattice model, 
however, there is an 
umklapp term which breaks scale invariance. The coupling constant of this term renormalizes 
to zero as the system is studied on increasingly large length scales. In other words, field 
theory results generally do not relate to all observables of the corresponding microscopic theory, 
but only to those measuring the system on certain length scales. The lower bound of this range 
of length scales is not sharply defined. Instead, understanding the {\it umklapp scale} $L_u$ as 
the typical length above which the {\it umklapp term is `too small to be noticed'} turns out to be 
convenient. Of course, there is no unique way to determine $L_u$. Different `measurements' and 
the corresponding definitions of what means `too small to be noticed' will give rise to 
different representations of $L_u$ (we indicate this by superscripts RG, O, and n in the 
following). However, they agree in their qualitative behavior, namely that $L_u$ is 
atomically small 
for $U/t \ll 1$ and diverges for $U/t \to 2$.

We shall first analyze the g-ology RG of the translational invariant model 
and identify the above-described effect in the RG flow. The umklapp term Eq.~\eqref{eq:H_um} 
in the bosonized theory with a coupling constant $y$ breaks the scale invariance, but 
is irrelevant in the RG sense for $U/t < 2$. In general, the RG produces a sequence of 
effective low-energy theories which are capable of describing the physics on increasingly 
large length scales. In this sequence $y$ decreases to zero and $K$ approaches $K_B$ 
as determined from the Bethe ansatz. However, an `exact RG', which is valid for all $y$, is 
not known. Instead one usually resorts to perturbative RG equations, valid for small umklapp 
amplitudes\cite{Giamarchi03}
\begin{align}
\frac{dy(l)}{dl} &= -y(l)[4K(l)-2], \nonumber \\
\frac{dK(l)}{dl} &= -y(l)^2K(l)^2 \label{flowequations}
\end{align}
where $L=e^l$ is the length scale above which the renormalized theory is valid. 
Clearly, for $\nicefrac{1}{2}<K<1$ (corresponding to $2 > U/t >0$), $y=0$ is a stable 
fixed point of the flow equations. The initial conditions for $K$ and $y$ can be 
obtained perturbatively from the lattice model, that is for $U/t \ll 1$. However, 
since we are more interested in the end of the RG flow than in its beginning, it is convenient to 
fix the end point $[K(\infty),y(\infty)] = [K_B,0]$ with the help of the Bethe 
ansatz solution of our translational invariant lattice model. 
Thus, the question is: `For which length scale $L_u = e^{l_u}$ is $y<y_0$ 
for a certain small $y_0$, which we are free to choose.'

\begin{figure}
\centering
\includegraphics[width=121pt]{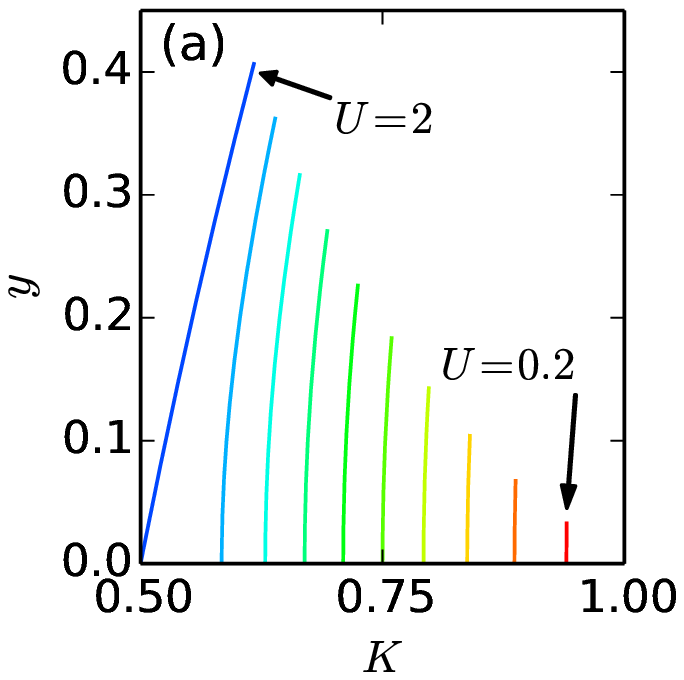}
\includegraphics[width=121pt]{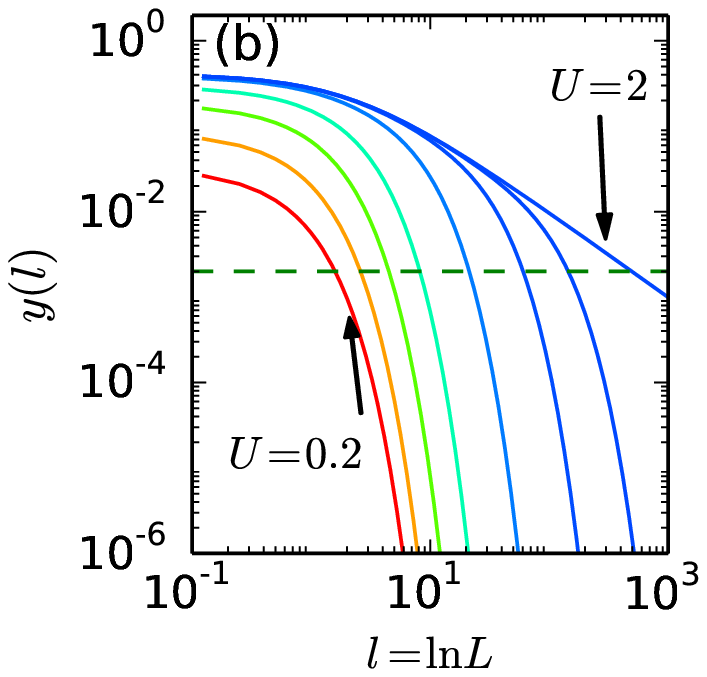}
\caption{(Color online) Perturbative RG flow for different $U$. Part (a) shows the flow in 
the $Ky$-plane for $U=0.2,0.4,\dots,2$. Part (b) shows $y(l)$ for 
$U=0.2,0.5,1,1.5,1.9,1.99,1.999,2$ on a log-log scale. The  dashed (green) line represents 
the $y_0$ we have chosen for determining $L_u^{RG}$.}
\label{fig_rgflow}
\end{figure}

\begin{figure}
\centering
\includegraphics[width=200pt]{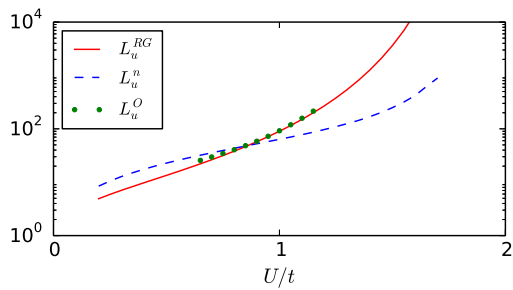}
\caption{(Color online) $U$-dependence of different representatives of the umklapp length scale $L_u$: The 
RG-based $L_u^{\rm RG}$, the $n(k)$-based $L_u^{\rm n}$, and the overlap-based $L_u^{\rm O}
$. }
\label{fig_Lu}
\end{figure}

Figure \ref{fig_rgflow} shows the solution of the approximate flow equations 
(\ref{flowequations}) for various $0<U/t\leq 2$. The RG flow always seems to reach its 
fixed point $[K_B,0]$, but Fig.~\ref{fig_rgflow}(b) makes clear that the flow equations must 
be integrated over increasingly large scales $l$ as $U/t$ gets close to its critical 
value 2. We may now define the weak coupling RG-based umklapp scale $L_u^{\rm RG}$ implicitly 
by 
\begin{equation}
y(\ln[L_u^{\rm RG}]) = y_0.
\end{equation}
We choose $y_0=0.002$ and plot $L_u^{\rm RG}$ as a function of $U/t$ in 
Fig.~\ref{fig_Lu} as a solid line. 
It is apparent that this representative of the umklapp length scale becomes exceedingly 
large as $U/t\to 2$.

\begin{figure}
\includegraphics[width=\linewidth]{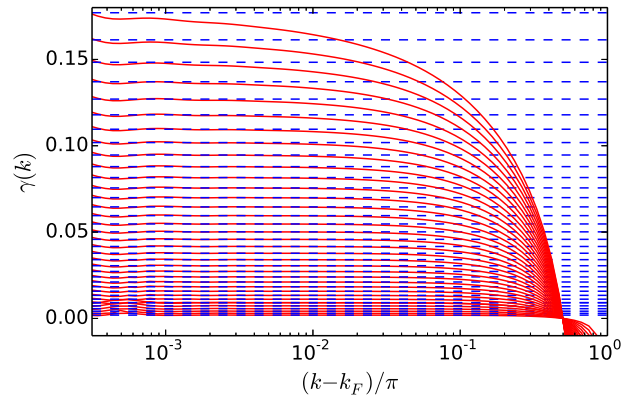}
\caption{(Color online) Exponents $\gamma$ of the momentum distribution function 
$n(k)$ for $U=0.2,0.25,\dots,1.95$ (from bottom to top). The dashed (blue) lines represent the 
expectations $\gamma_B = (K_B +1/K_B-2)/2$ from the Bethe ansatz. The solid red lines are the 
logarithmic derivatives $\gamma(k)$ (see Eq. (\ref{nk_logder})).}
\label{fig_nk_fan}
\end{figure}

A second way of determining $L_u$ is based on the power-law behavior of the 
momentum distribution function $n(k) = \langle c^\dagger_k c_k\rangle$ for $k\approx k_F$. 
From bosonization it is known that 
\begin{equation}
|n(k)-1/2| \sim |k-k_F|^{\gamma_B} \label{nk_powerlaw},
\end{equation}
with $\gamma_B = (K_B +1/K_B-2)/2 $. The usage of the Bethe-ansatz-based $K_B$ in the 
exponent 
indicates that it has been assumed implicitly that $y$ has been renormalized to zero already. 
Indeed, the derivation of Eq.~(\ref{nk_powerlaw}) within the TLM requires the absence of the 
umklapp term. As a consequence, this power law should be detectable only on length scales 
beyond 
$L_u$. This means that, in a numerical simulation of the lattice model, the range of validity 
$|k-k_F| \lesssim 1/L_u$ of the power law decreases as $U$ increases.

This can be observed in numerical simulations of the lattice model. 
Figure \ref{fig_nk_fan} shows the logarithmic derivatives of $1/2-n(k)$
(for $k-k_F >0$)
\begin{equation}
\gamma(k) = \frac{d \ln [1/2-n(k)]}{d \ln (k-k_F)}\label{nk_logder}
\end{equation}
as function of $\log_{10} (k-k_F)$, extracted from an iDMRG ground state calculation with 
bond dimension 1600 (for details, see Appendix \ref{ap_DMRG}). 
The dashed lines show the asymptotic expectation $\gamma_B$ for $\gamma(k)$ based on the 
Bethe ansatz. Apparently, 
the range in which the numerical results agree with the Bethe-ansatz expectations becomes 
smaller 
as $U$ grows. This was earlier found in similar calculations but not analyzed in 
detail.\cite{Karrasch12} 
In analogy with the RG-based umklapp scale, we may define the $n(k)$-based 
umklapp scale $L^{\rm n}_u$ via
\begin{equation}
|\gamma(k_F + \pi/L_u^{\rm n}) - \gamma_B| < \delta,
\end{equation}
where the small $\delta$ is the difference between $\gamma_B$ and the numerical 
$\gamma(k)$ we are prepared to accept. For the data of Fig.~\ref{fig_Lu} we chose $\delta=0.001$ and 
observe qualitatively the same behavior as for the other $L_u$-representatives.

A third way to determine $L_u$ is based on fitting the CFT form Eq.~(\ref{fitform}) to 
{\em overlaps} 
between ground states of the lattice model of size $L$ with (infinite) bond impurity $b=0$ and without 
the latter (OBC; see Fig.~\ref{fig_b0_logform}). If one restricts the $L$-range of 
the data to be fitted to $[L_r-\Delta L, L_r +\Delta L]$, one observes convergence of the 
extrapolated exponent $\alpha$ as the range is shifted towards larger $L_r$. From the 
systematic investigation of how large systems are needed in order to predict the correct 
asymptotic exponent $\alpha=1/16$, one may extract the overlap-based umklapp scale $L_u^{\rm 
O}$.
Details about this procedure can be found in Appendix \ref{vrf_Lu}.

In Fig.~\ref{fig_Lu} we compare the three length scales $L_u^{\rm RG/n/O}$, which are all 
representatives of the same physical effect, namely that the lattice model's correspondence 
with a CFT requires exceedingly large length scales as $U/t$ is increased towards its critical value 2. 
It is not surprising that the representatives differ significantly from each other in their 
detailed form, since the criterion `too small to be noticed' is not directly comparable for the 
different aspects discussed above. In fact, the excellent agreement of $L^{\rm O}_u$ and 
$L_u^{\rm RG}$ should be viewed as a coincidence. The essential feature shared by all the related 
length scales is that they are atomically small for $U\approx 0$ and grow exponentially for $U/t 
\gtrsim 1$.
As exemplified considering $n(k)$ typical LL power laws can only be expected on momentum scales 
smaller than $L_u^{-1}$ (see also Ref.~\onlinecite{Karrasch12}). While the appearance of such an 
interaction dependent scale associated to RG irrelevant two-particle scattering terms is routinely 
considered in spinful lattice models such as the 1D Hubbard model (flow of $g_{1,\perp}$-term 
in the g-ology classification) its role was so far not fully appreciated in studies of the 
spinless lattice model with nearest-neighbor interaction.

\begin{figure}[h]
\includegraphics[width=0.9\linewidth]{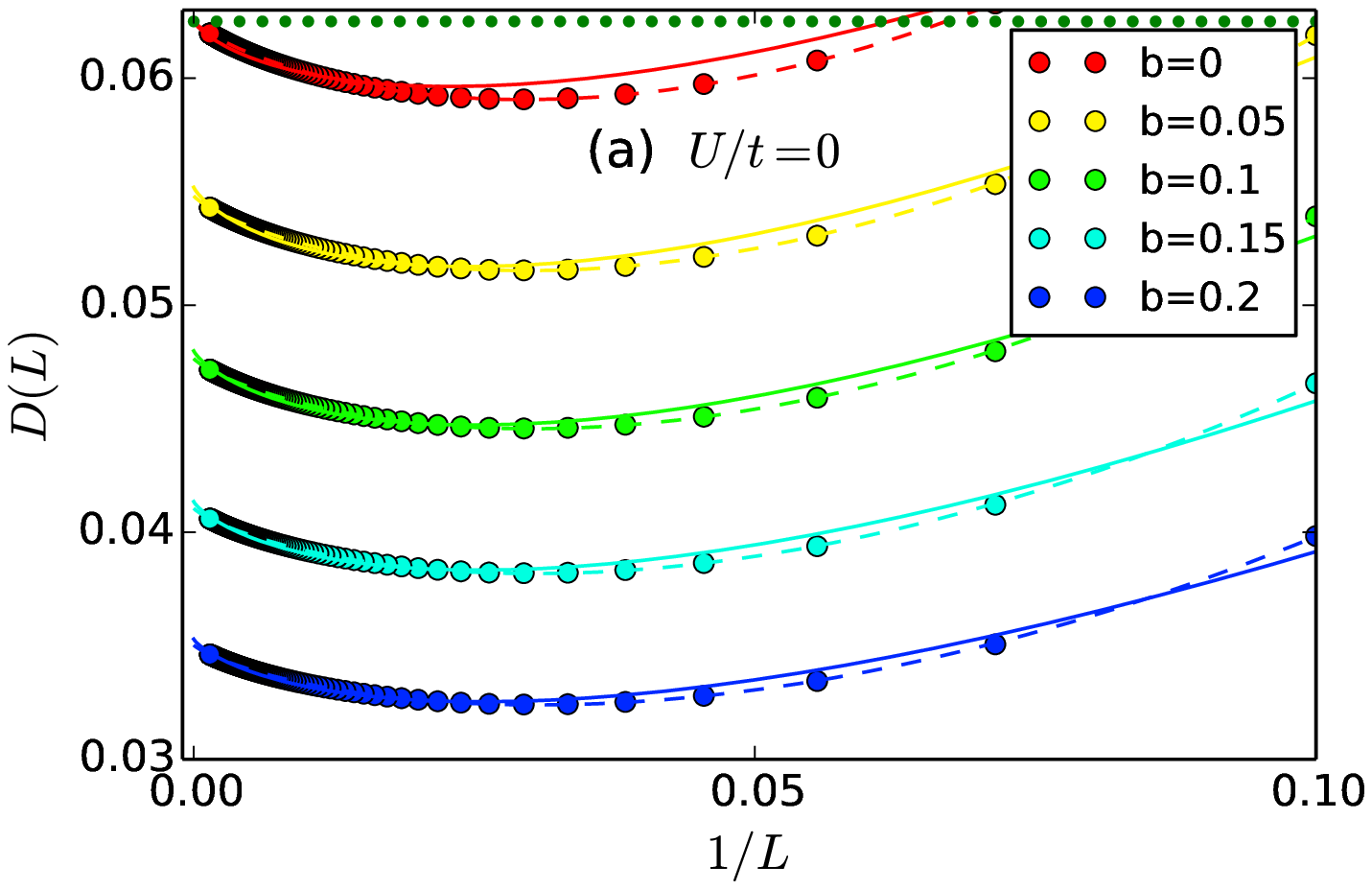}
\includegraphics[width=0.9\linewidth]{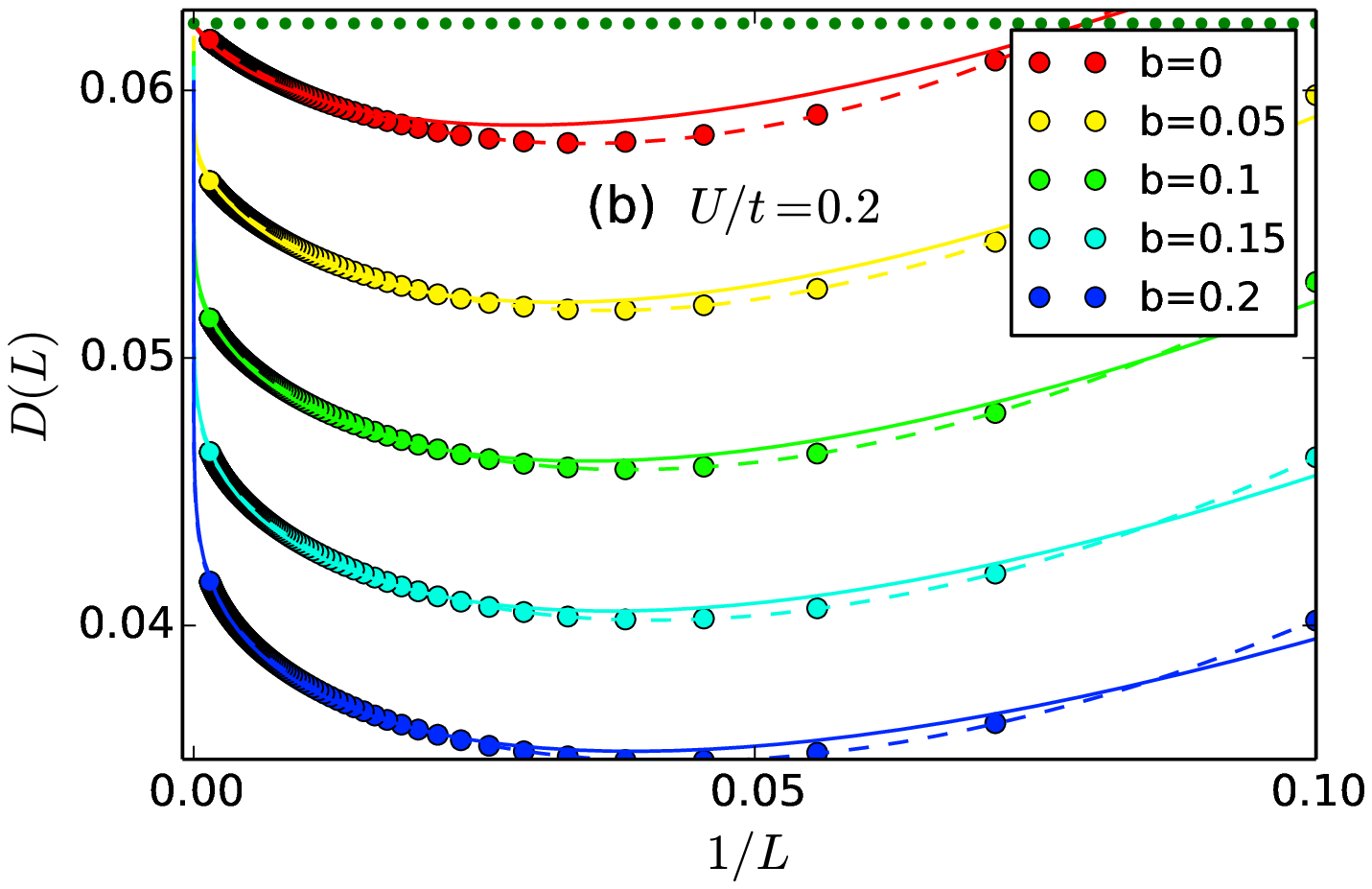}
\includegraphics[width=0.9\linewidth]{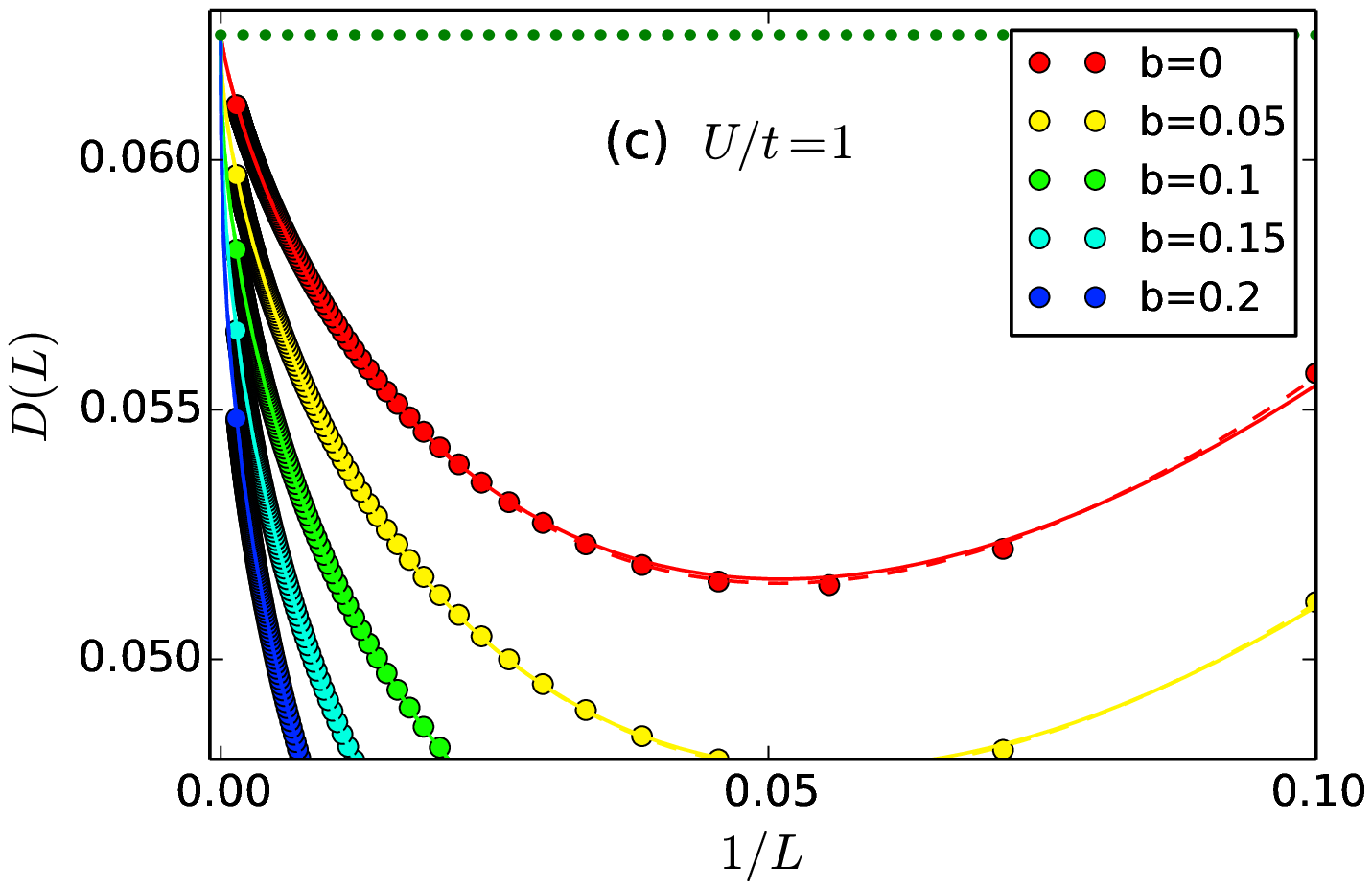}
\caption[]{(Color online) Logarithmic derivatives of the numerical overlaps 
$\langle \OBC{baseline=height} | \OBCb{baseline=height}
\rangle$
(dots) and the best fits (full lines) of the form  Eq. (\ref{fitform_kf})
for systems with up to $L=2000$ lattice sites. 
For $U=0$ the curves extrapolate to the noninteracting exponents
Eq.~(\ref{eq:OE0}). For $U>0$ all curves 
extrapolate to 1/16 for $L\rightarrow\infty$. The dotted (green) horizontal line indicates 
the limiting exponent 1/16. The dashed lines show fits with an additional $L^{-2}$ term.
Note the different $y$-axis scales of (a)-(c).}
\label{fig_obc_b}
\end{figure}

For our further analysis this means that one should always keep in mind that a minimum system 
size 
is required if CFT arguments are to be used, and that this minimum system size grows very 
strongly for 
$U/t \gtrsim 1$.

\section{Bond impurities}

\label{sect_weak_link}

\begin{figure}[h]
\includegraphics[width=0.9\linewidth]{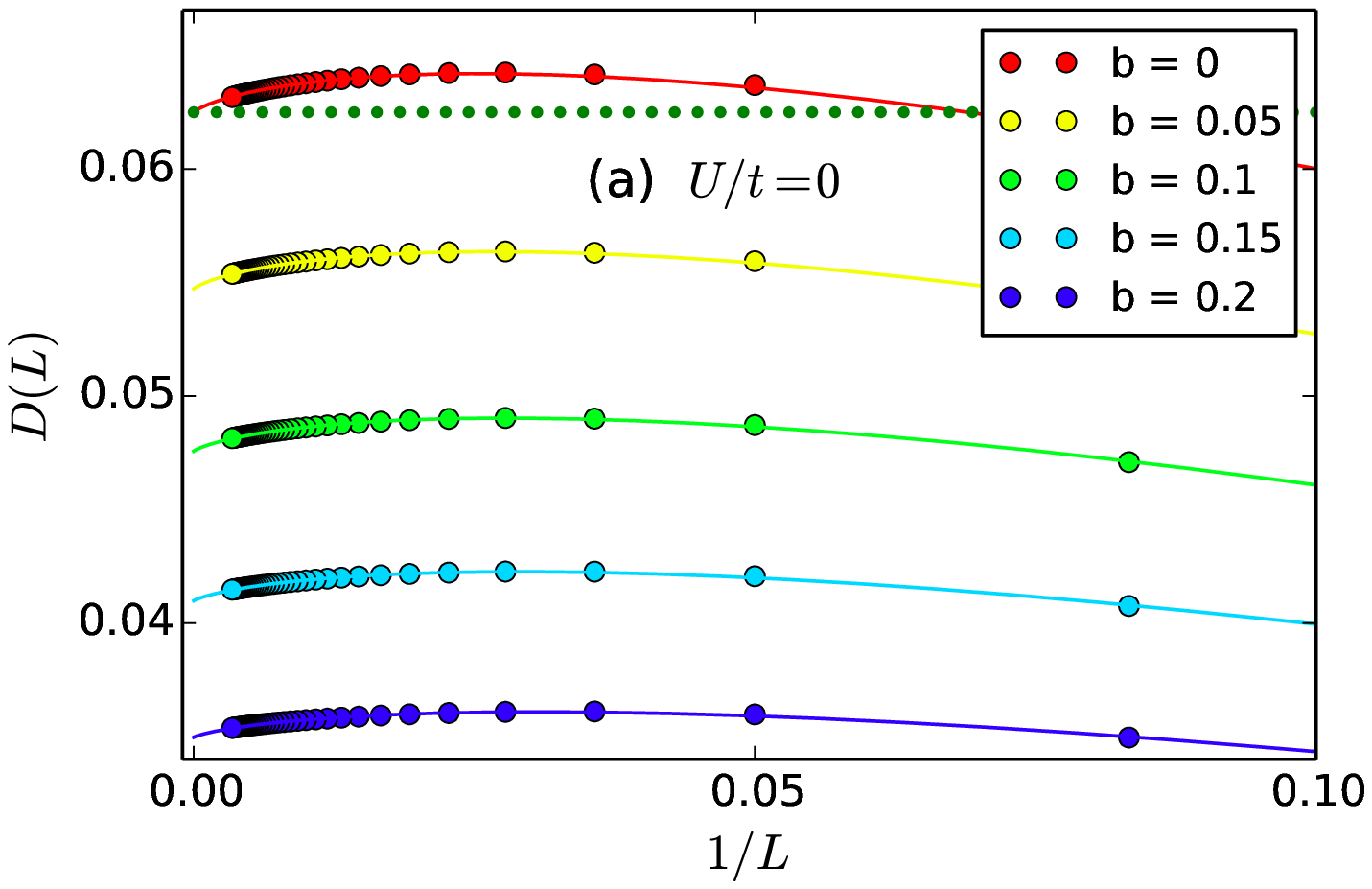}
\includegraphics[width=0.9\linewidth]{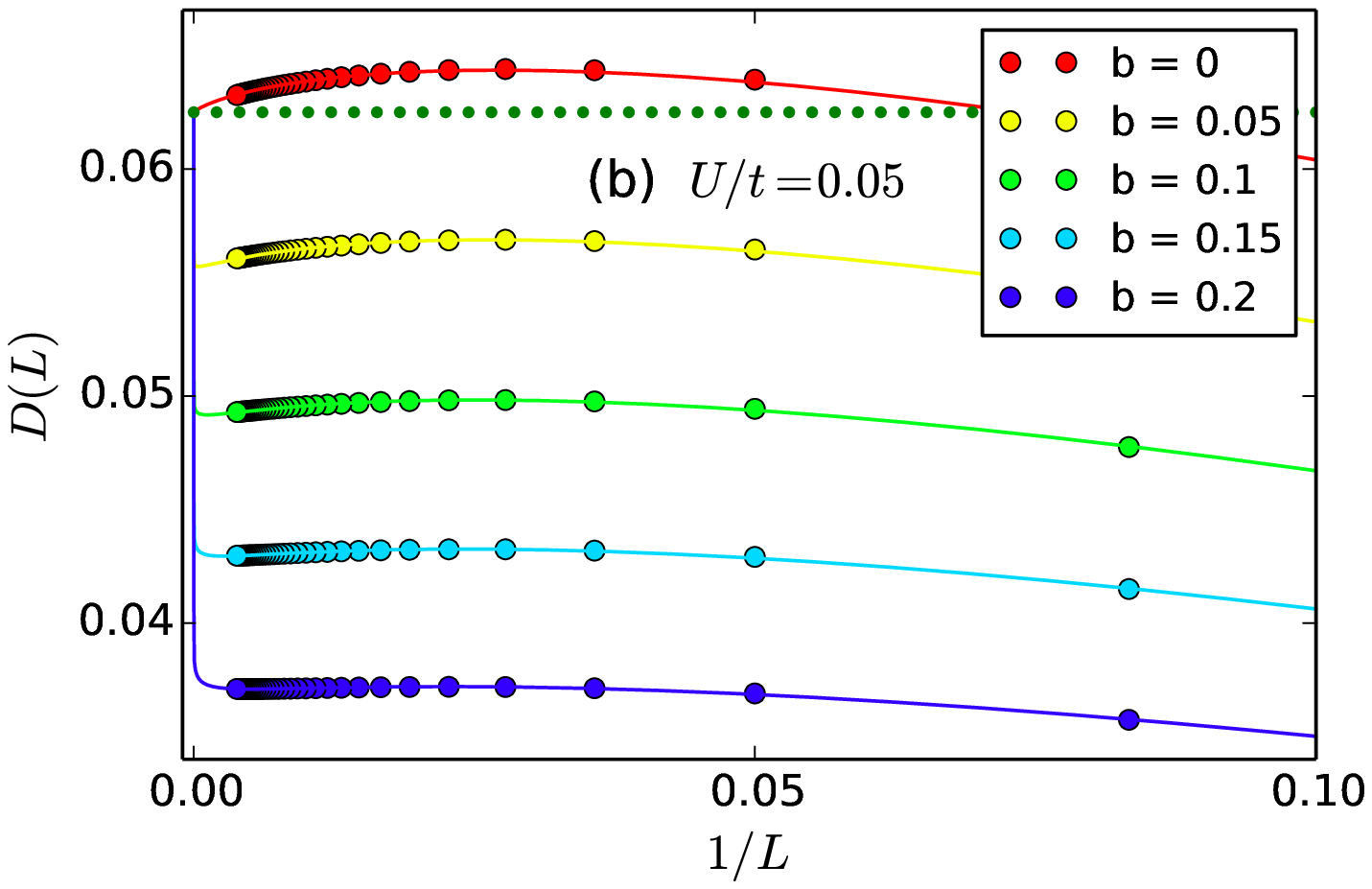}
\includegraphics[width=0.9\linewidth]{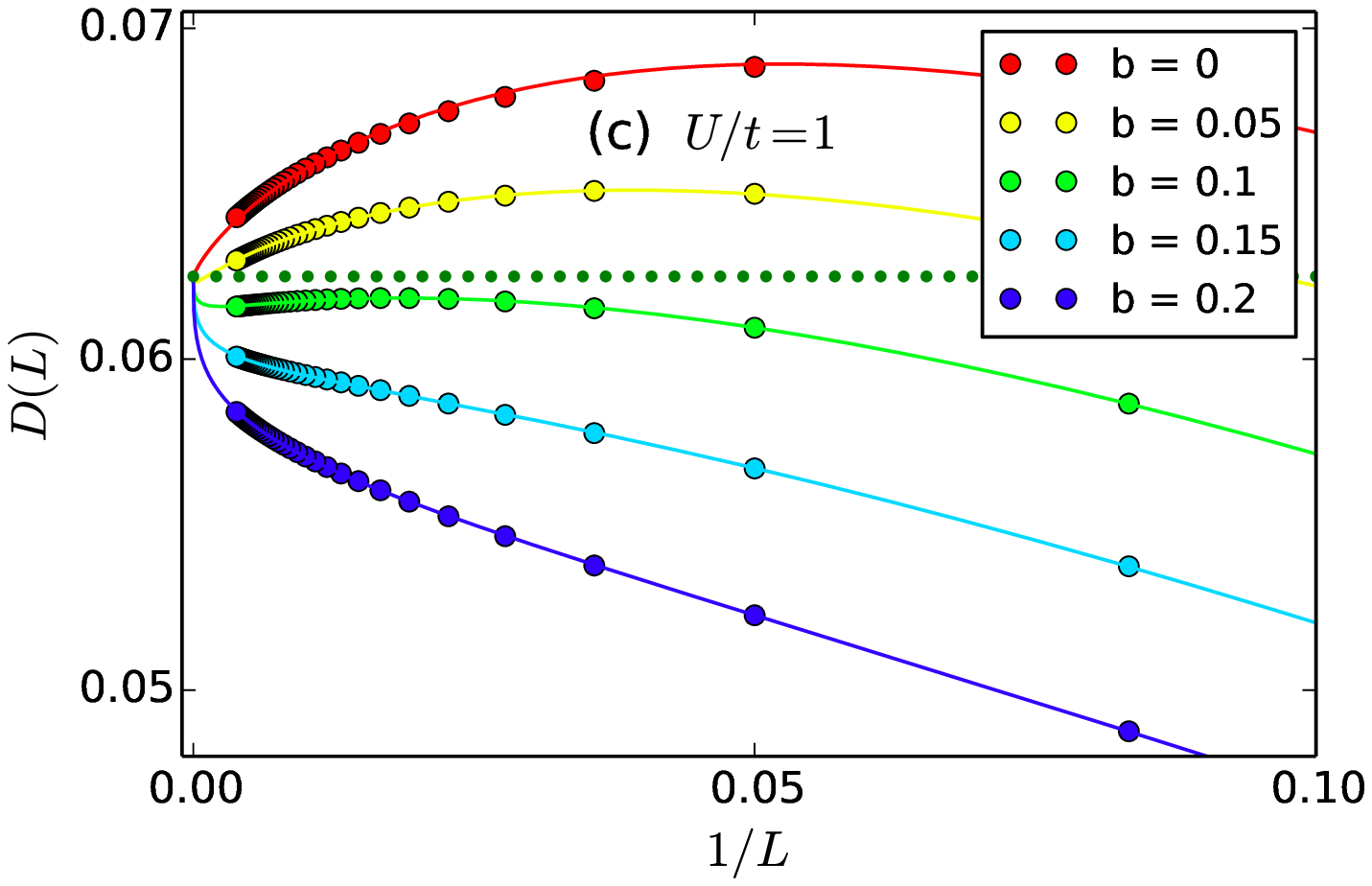}
\caption[]{(Color online) Logarithmic derivatives of the numerical overlaps 
$\langle \PBC{baseline=height} | \PBCb{baseline=height}
\rangle$ (dots) and the best fits (lines) of the form  
Eq. (\ref{fitform_kf}) for systems with up to $L=256$ lattice sites. 
For $U=0$ the curves extrapolate to the noninteracting exponents Eq.~\eqref{eq:OE0}. For $U>0$ all curves extrapolate to 1/16 for 
$L\rightarrow\infty$. The dotted (green) horizontal line indicates the limiting exponent 1/16.}
\label{fig_pbc_b}
\end{figure}

We are now prepared to investigate the OC in our lattice model with finite bond impurity 
$0< b <1$ considering OBC (overlap $\langle \OBC{baseline=height} | \OBCblone{baseline=height}{0}
\rangle$) as 
well as PBC (overlap $\langle \PBC{baseline=height} | \PBCblone{baseline=height}{0}
\rangle$). 
As before, we study the length dependence of the discrete logarithmic derivatives of ground state 
overlaps $D(L)$ Eq. (\ref{discrete_log_der}) and now fit the form Eq.~\eqref{fitform_kf} to those. 
It is the combination of three `subleading effects' that governs the approach to the 
large-$L$ limit: (i) the systems must be longer than the umklapp scale $L_u$ which
we ensure by considering $U/t \lessapprox 1$ only. (ii) The 
stress-tensor-based subleading correction $\sim\ln (L)/L$ must be respected. (iii) 
The impurity scaling term $\sim L^{1-1/K}$ must be included, as well. 
In our fits for finite $b$ and $U>0$, we {\em fix} $\alpha=1/16$ and $K=K_B$, which leaves us 
with the three fit parameters $\beta,\lambda$, and $\kappa$. For $U=0$, the exponent of 
the impurity scaling term is zero and we may absorb $\kappa$ into $\alpha$, thus 
leaving the OE exponent as a fit parameter in this noninteracting limit as well.

\begin{figure}
\includegraphics[width=\linewidth]{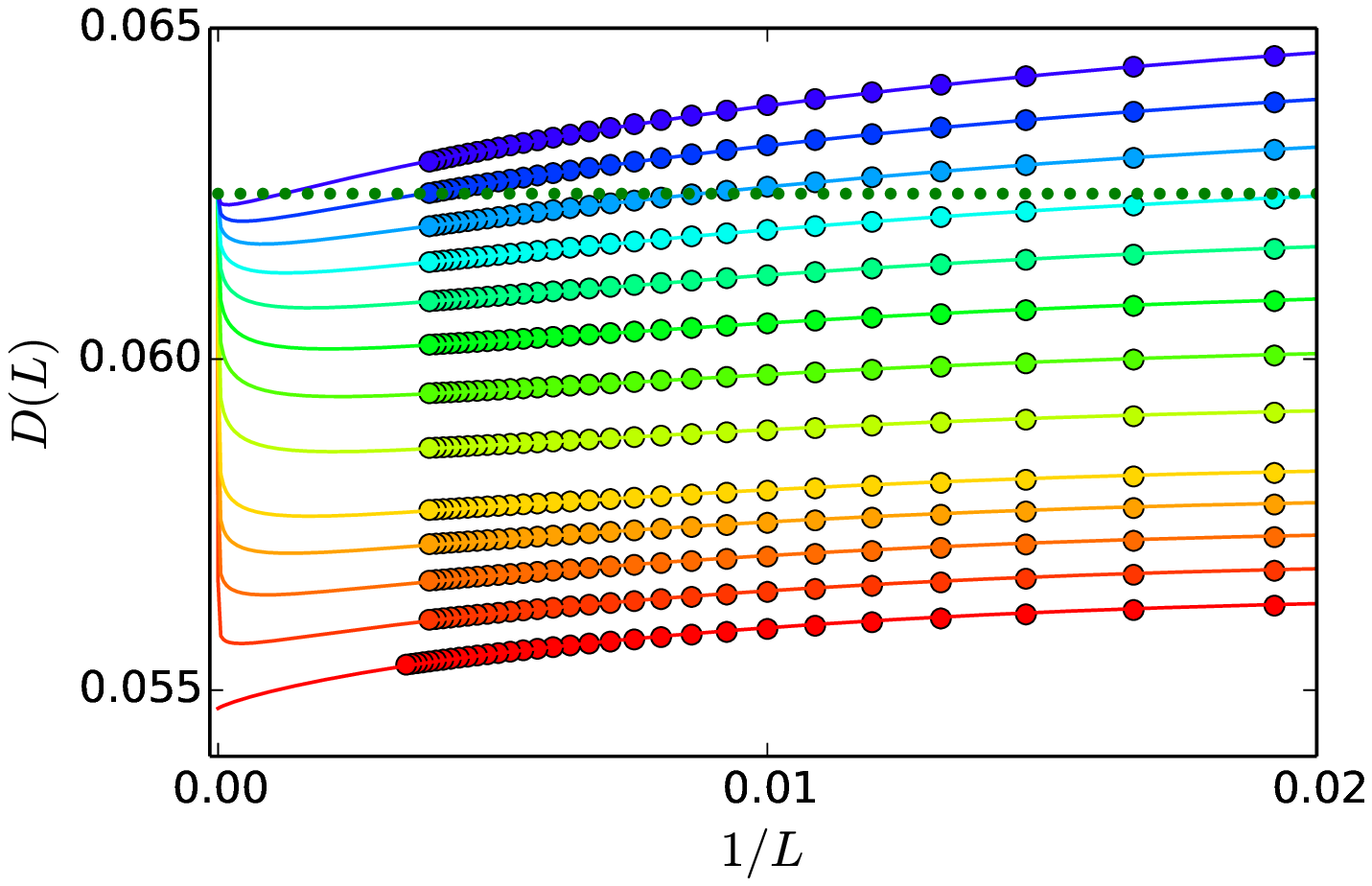}
\caption[]{(Color online) Logarithmic derivatives of the numerical 
overlaps $\langle \PBC{baseline=height} | \PBCbvthree{baseline=height}{0.05}
\rangle$ for $b=0.05$ (dots) and the best fits (lines) to the form  
Eq. (\ref{fitform_kf}) for systems with up to $L=256$ lattice sites. The interaction strength is  
$U/t=0,0.05,0.1,0.15,0.2,0.3,\dots,1$ from bottom to top (red to blue). All fit curves but 
the one for $U=0$ extrapolate to $1/16$ for $L\rightarrow\infty$. Note the nonmonotonic 
behavior. The dotted (green) horizontal line indicates the limiting exponent 1/16.}
\label{fig_b0p05_allV}
\end{figure}

We start out with OBC as those were also considered in the $b=0$ field theory 
studies Refs.~\onlinecite{Dubail11,Stephan11,Stephan13} and analyze the overlap 
$\langle \OBC{baseline=height} | \OBCblone{baseline=height}{0}
\rangle$. 
Figure \ref{fig_obc_b} shows the DMRG-based logarithmic derivatives of the 
overlaps $D(L)$ as a function of $1/L$ (circles) together with fits of form 
Eq.~(\ref{fitform_kf}) (solid lines) for systems of up to $L=2000$. For $U=0$ the curves extrapolate 
to the 
noninteracting finite-$b$ limits of the OE $\alpha_{U=0} < 1/16$ given 
in Eq.~(\ref{eq:OE0}), as expected  [Fig.~\ref{fig_obc_b}(a)]. 
Since $K=1$ in this case, the impurity scaling term is not effective. 
For $U > 0$, however, it leads to a severe finite-size correction. For 
$U/t=0.2$ [Fig.~\ref{fig_obc_b}(b)], that is, $K_B\approx 0.940$, where the exponent of the impurity flow is 
$1-1/K_B \approx -0.0638$. As a consequence of this slowly decaying finite-size 
correction and the (smaller) $\ln(L)/L$ correction, the asymptotic regime, where $D(L)$ is significantly closer to 1/16 than to $\alpha_{U=0}$, is virtually 
never reached. Without the knowledge of these extreme subleading terms and only on the 
basis of the numerical overlaps for up to a few thousand lattice sites, one would never 
be able to properly perform the extrapolation in Fig.~\ref{fig_obc_b}(b) to the 
asymptotic regime where $D(L) \approx \alpha =1/16$. The situation improves 
for $U/t \approx 1$ [Fig.~\ref{fig_obc_b}(c)] and the data are closer to the asymptotic value. Note the different $y$-axis scales 
of Fig.~\ref{fig_obc_b}(a)-(c). 
The finite-size fits for open boundary conditions in Fig. \ref{fig_obc_b} to the form with $L^{-1}$ 
as the most quickly decaying term show slight deviations from the numerical data for 
small $L$ (solid lines in Fig.~\ref{fig_obc_b}). Including a further term 
$\sim L^{-2}$ in the functional form improves the fit quality considerably (dashed lines
in Fig.~\ref{fig_obc_b}).

Even though for $U > 0$ we fixed $\alpha$ to $1/16$ we judge the excellent agreement between
the DMRG data and the fits to provide {\em strong evidence} for the asymptotic value 
$1/16$ of the backscattering component of the OE in our lattice model. Furthermore, the 
quality of the fits gives us confidence that the scaling form Eq.~\eqref{fitform_kf} which 
was based on field theoretical arguments indeed presents the leading finite size corrections of 
$D$ of the microscopic model. 

Figures \ref{fig_pbc_b} and \ref{fig_b0p05_allV} show the fits for the overlap 
$O=\langle \PBC{baseline=height} | \PBCb{baseline=height}
\rangle$ and system sizes of up to $L=264$. It is inherent to the
DMRG algorithm that for PBC the numerical resources are exhausted faster than for OBC which 
explains 
the difference in reachable system sizes.
For $U=0$ the numerical data again extrapolates to the known noninteracting OE 
Eq.~(\ref{eq:OE0}). 
For $U>0$ our conclusions are identical to the case of OBC.  
We note, however, that the quality of the fits to Eq. (\ref{fitform_kf}) without an additional 
$L^{-2}$ term is as good as the one of the fits for OBC including this term. Since finite size
effects are expected to be more severe for OBC than for PBC, 
this is not surprising. The excellent quality of the fits provides evidence that the CFT 
arguments, employed for OBC, which led us to consider the form Eq.~(\ref{fitform_kf}) of the 
finite size corrections are applicable to both, OBC as well as PBC. Note, that for $U=0$ and $b=0$ 
this was already hinted at in Ref.~\onlinecite{Stephan13}. 

The excellent agreement between the DMRG data for $D(L)$ and the fits by the form 
Eq.~(\ref{fitform_kf}) for OBC and PBC is naturally linked to our conclusion that the 
backscattering contribution to the OE in the microscopic model is indeed given by $1/16$. 
This consistency of the numerical data and the expected analytical finite-size corrections constitutes the third important result of our present work.  

It is worth noting that the prefactors of the $\ln(L)/L$ terms have opposite signs 
for different boundary conditions, while the impurity scaling term has equal signs. Since these 
are the most slowly decaying finite size contributions, this sign difference leads to counterintuitive 
behavior of the logarithmic derivatives for PBC: as shown in 
Fig. \ref{fig_pbc_b} and in particular Fig. \ref{fig_b0p05_allV}, for certain parameter 
combinations $U/t$ and $b$, $D(L)$ seems to decrease 
monotonously and even appears to converge, until for system sizes beyond the reach of 
numerical techniques it turns up and finally approaches the asymptotic regime with $\alpha=1/16$. 
This explains why the results of the earlier DMRG studies performed for PBC turned out to be 
inconclusive.\cite{Qin96,Qin97,Meden98}

\section{Summary}

\label{sect_sum}

Using state of the art DMRG we have studied the orthogonality catastrophe in the lattice model of spinless fermions with 
repulsive
nearest-neighbor interaction at half band filling. We were able to provide convincing evidence 
for the expected backscattering contribution $1/16$ to the asymptotic orthogonality exponent for weak to intermediate 
interactions. This was only possible by carefully considering finite size corrections stemming from
the field theoretical insight that the logarithm of the overlap can be viewed as a free energy. For
chains with periodic boundary conditions the interplay of these terms results in nonmonotonic scaling behavior for system 
sizes 
$L \to \infty$ (compare also Fig.~5 of Ref.~\onlinecite{Meden98}). For interactions approaching the one at which umklapp scattering becomes RG 
relevant and 
the system enters a charge-density wave phase we were not able to confirm $\alpha \to 1/16$. For 
such 
results from scale invariant field theory cannot be employed on the reachable length 
scales of up to a few thousand lattice sites due to residual umklapp scattering. 

For the overlap between ground states of a homogeneous and a not perfectly cut system ($b>0$) the true asymptotic regime for the orthogonality exponent, where all subleading terms are negligibly small, is virtually unreachable for small interactions $U$. It is important to note that the system sizes needed for reaching the asymptotic regime not only exceed the computational resources, but are also beyond experimental reach. For instance, at $U/t=0.1$ and $b=0.05$ (compare also Fig. \ref{fig_b0p05_allV}), at $L=10^{9}$ lattice sites, corresponding to system sizes in the meter range, the observed exponent 0.058 is still closer to $\alpha_{U=0}\approx 0.055$ than to the asymptotic $1/16=0.0625$. Moreover, for reasonable system sizes the exponent {\it seems to converge} to a value less than $1/16$.

We expect to find similar behavior in other half-filled lattice models.
Our study highlights the importance of scaling of single-particle inhomogeneities as well as 
two-particle scattering in 1D correlated electron systems in their Luttinger liquid phase.  

\section*{Acknowledgments}
We are grateful to J\'er\^{o}me Dubail, Christoph Karrasch, Ian McCulloch, and Romain Vasseur for 
enlightening discussions. This 
work was 
supported by the DFG via Research Training Group 1995 `Quantum many-body methods in 
condensed matter systems'.

\appendix

\section{Variable window fits of overlaps\label{vrf_Lu}}

\label{ap_variable}

For investigating the asymptotic behavior of a quantity and especially for extracting this behavior 
from numerical data, it is not always useful to use the full range of data points at hand for one single fit. 
Instead, one may restrict the range of fit data to a certain window of all available data and shift 
this window up and down. From such a procedure, one can learn something about the stability of 
such 
a fit. Moreover, effects which are not accounted for in the fit function, but are only present in 
a certain regime, can be identified. It is the latter aspect we are interested in, in the present context.

In this procedure, which we call {\it variable window fitting}, we restrict the numerical data  
for a particular fit to the window $[L_w-\Delta L/2, L_w +\Delta L/2]$. The size of the window 
$\Delta L$ is fixed and must be sufficiently large for the individual fits to be numerically 
stable. The best fit parameters can then be studied as functions of the center of the data 
window $L_w$. Here we are especially interested in the $L_w$-dependence of the extrapolated 
exponent $\alpha(L_w)$. 

\begin{figure}
\centering
\includegraphics[width=\linewidth]{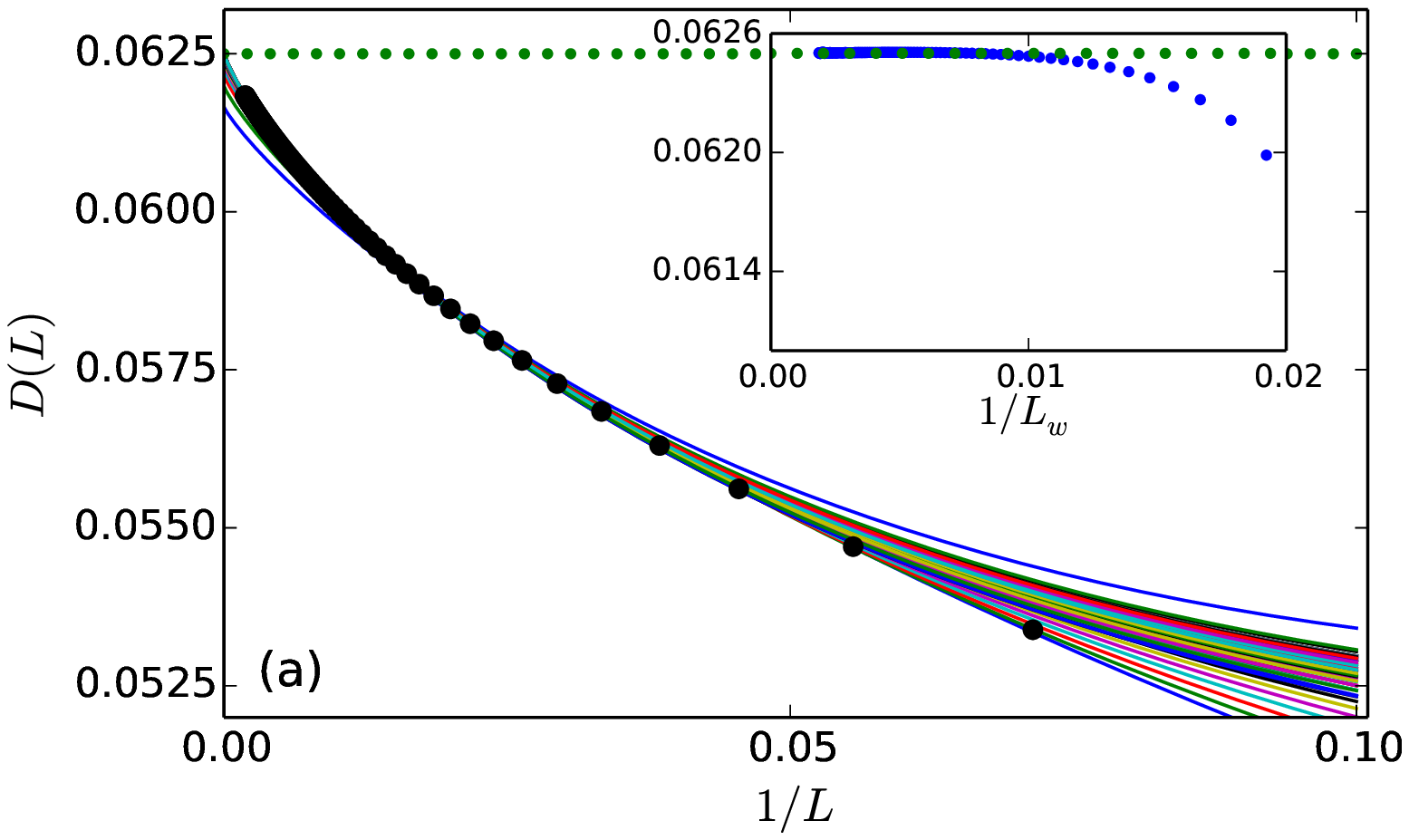}
\includegraphics[width=\linewidth]{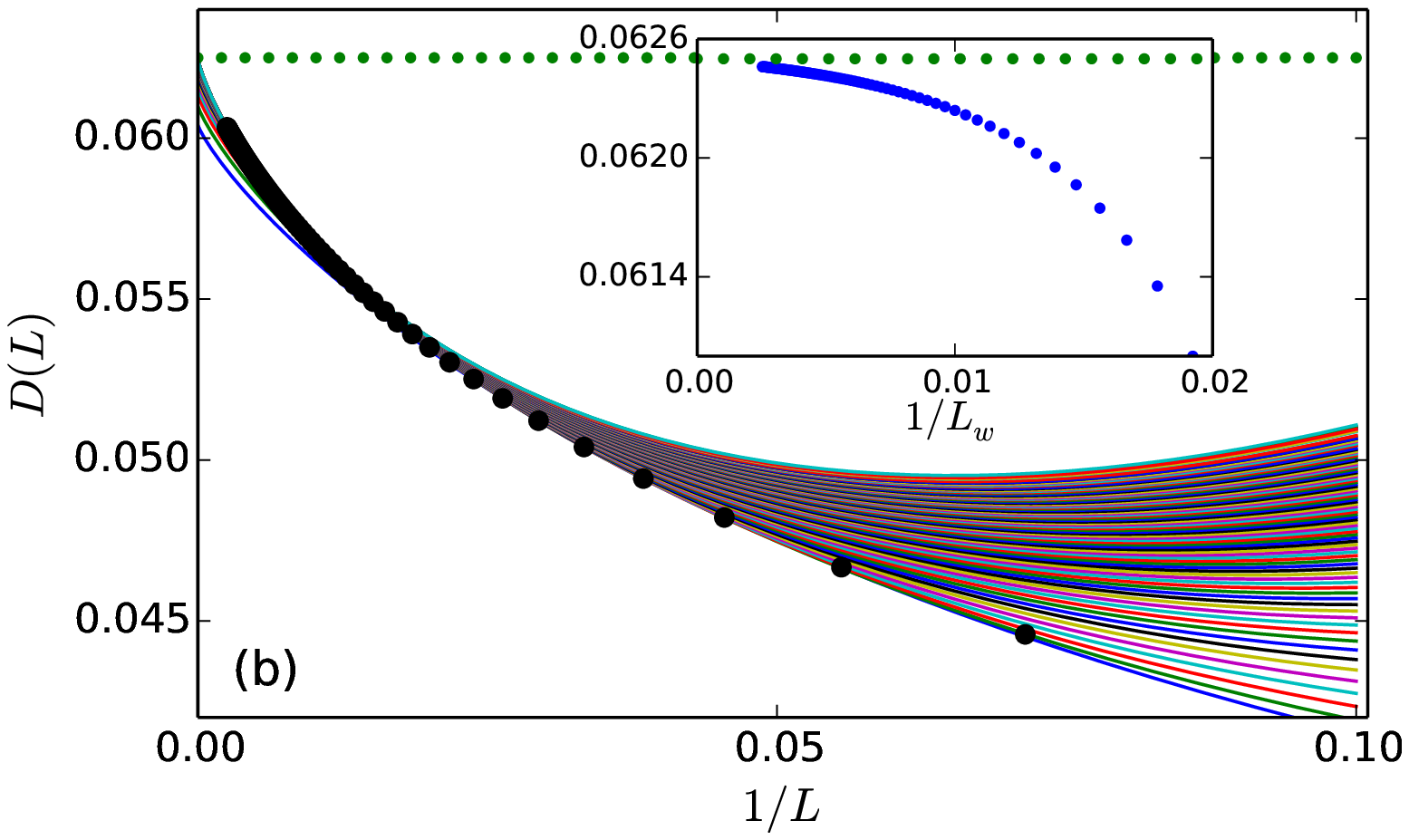}
\caption[]{(Color online) Variable window fits of the logarithmic derivatives of the overlap $O=\langle \OBC{baseline=height} | \OBCbvone{baseline=height}{0}
\rangle$: (a) for $U=0$ and (b) for $U/t=1$. The insets show the extrapolated 
exponents 
as a function of the inverse fit-window center $L_w^{-1}$.  The dotted (green) horizontal line indicates 
the limiting exponent 1/16.}
\label{fig_Lu_obc}
\end{figure}

\begin{figure}
\centering
\includegraphics[width=0.9\linewidth]{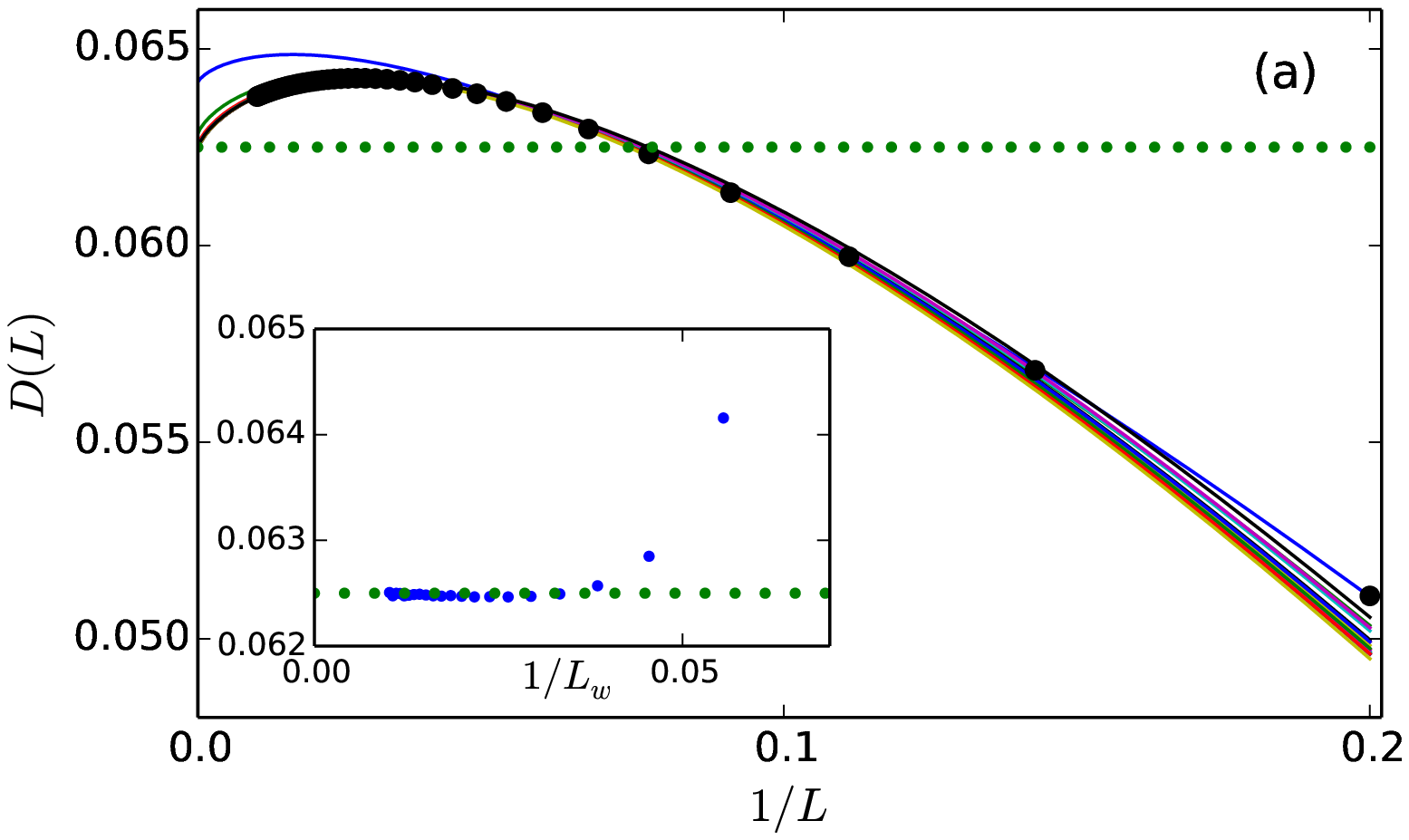}
\includegraphics[width=0.9\linewidth]{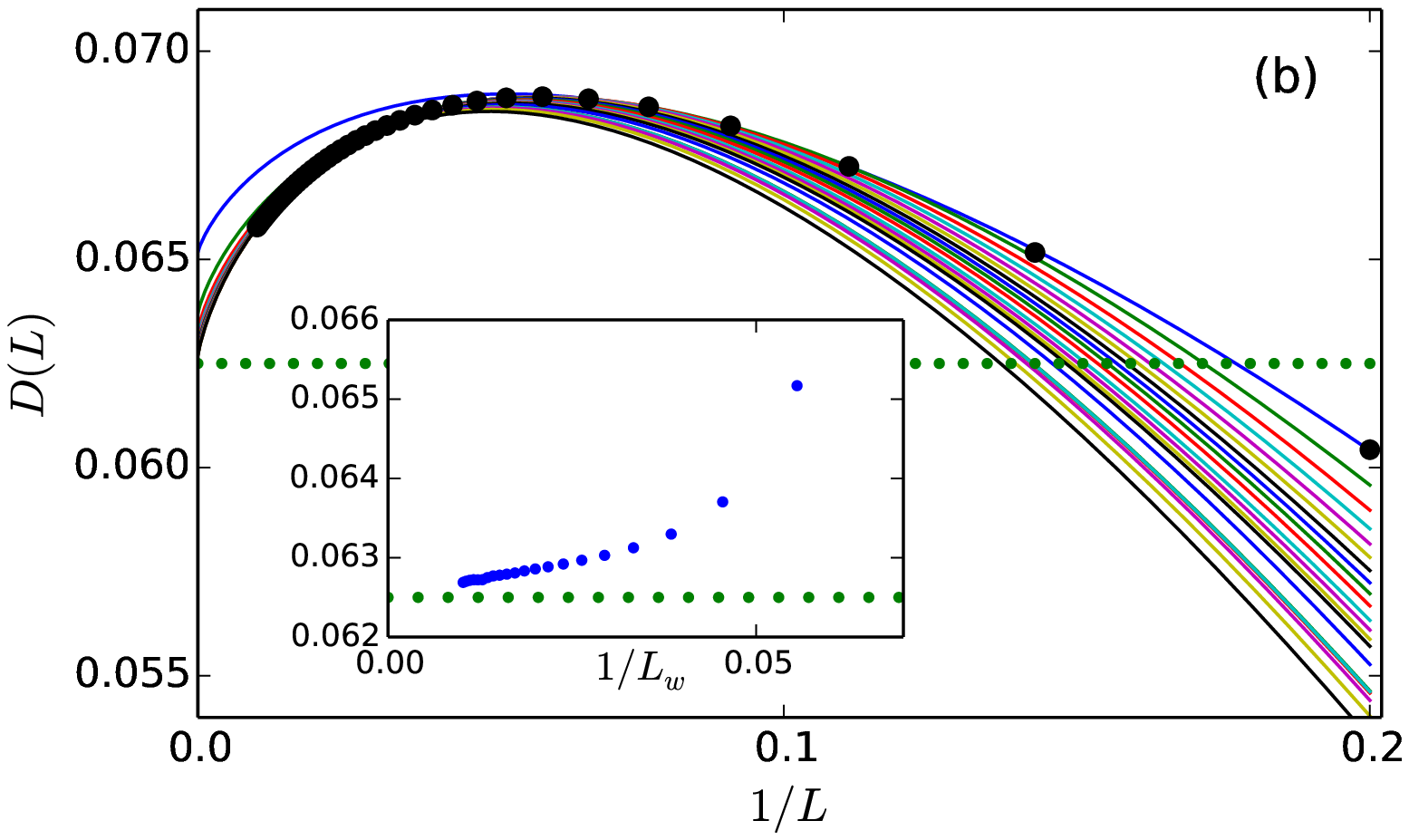}
\caption[]{(Color online) Variable window fits of the logarithmic derivatives of the overlap $O=\langle \PBC{baseline=height} | \PBCbvone{baseline=height}{0}
\rangle$: (a) for $U=0$ and (b) for $U/t=1$. The insets show the extrapolated 
exponents as 
a function of the inverse fit-window center $L_w^{-1}$.  The dotted (green) horizontal line indicates 
the limiting exponent 1/16.}
\label{fig_Lu_pbc}
\end{figure}

For overlaps with $b=0$ impurities the impurity scaling is irrelevant and thus the form 
Eq.~(\ref{fitform}) should be used for fitting the discrete logarithmic derivatives of 
the numerically calculated overlaps [see Eq. (\ref{discrete_log_der})]. However, since 
Eq.~(\ref{fitform}) originates from a CFT analysis, the umklapp process [described by Eq. (\ref{eq:H_um})] present in the 
interacting lattice model Eq.~(\ref{eq:Lattice_Ham}) with $U > 0$, which is the basis of 
the numerical simulation, is not accounted for in this fit form. As the length scales 
on which the system is studied is increased, the coupling constant $y$ of this umklapp process 
scales to zero. Thus, if such an umklapp term is present, one expects that the fit is 
only stable for sufficiently large $L_w$.

Figures \ref{fig_Lu_obc} and \ref{fig_Lu_pbc} show the variable window fits for $U=0$ and $1$ in 
systems with OBC and PBC, respectively. The different curves in the main plot correspond to 
different window centers $L_w$. We have chosen $\Delta L=40$ for the OBC and $\Delta L=14$ for the PBC data. Two effects leading to different 
curves for different $L_w$ should be distinguished. One is due to the quickly decaying contribution of the standard 
higher order finite-size corrections, such as $L^{-n}$ terms with $n \in {\mathbb N} $ and $n\geq 2$, which in the present case might be supplemented by terms $\ln(L)/L^n$.\cite{Dubail11,Stephan11,Stephan13} 
In the $U=0$ plots [parts (a) of Figs. \ref{fig_Lu_obc} and \ref{fig_Lu_pbc}], due to the absence of 
the umklapp term in the noninteracting limit, this is the only effect that can be observed. 
In this case, the convergence to $\alpha(L_w)\approx 1/16$ is reached already for small $L_w$, 
as can be seen in the corresponding insets. The second reason for deviating fits is the presence 
of terms in the Hamiltonian which are not captured by the scaling form Eq.~(\ref{fitform}). 
This effect can be observed in parts (b) of Figs. \ref{fig_Lu_obc} and \ref{fig_Lu_pbc}, showing 
the $U/t=1$ data with their corresponding variable window fits. As can be seen in the corresponding 
insets, the typical system sizes needed to achieve convergence to $\alpha=1/16$ are much larger 
than 
in the $U=0$ case. We attribute this to the umklapp length scale $L_u$ which diverges as $U/t$ 
approaches its critical value $2$.

From the variable window fits we may extract a further representative of $L_u$, namely 
the overlap-based umklapp scale $L_u^{\rm O}$. For this we fix an acceptance interval of a certain 
width $\Delta\alpha$ around $\alpha=1/16$. If the extrapolation $\alpha(L_w)$ from a certain 
data window $L_w$ is in this interval, then the $L_w > L_u^{\rm O}$. Thus, $L_u^{\rm O}$ is 
defined as 
the $L_w$ for which the curve $\alpha(L_w)$ enters the acceptance interval. We have 
chosen $\Delta\alpha=0.0002$ as the interval width from which the $L_u^{\rm O}$ in Fig. 
\ref{fig_Lu} 
has been extracted.

\section{Details about the DMRG calculations}

\label{ap_DMRG}

All numerical results discussed in this work have been acquired by the DMRG method. We have employed two 
versions 
of the general DMRG concept.\cite{Schollwoeck11} 

For the ground states of the lattice models with finite size $L$ we have used the standard iterative 
ground state finder with a two site update, formulated in matrix product state (MPS) language. 
In order to increase the efficiency of our code, we have explicitly used the conservation of 
the total particle number. The method is limited by the discarded terms in the wave function. 
After each two site optimization, the wave function is Schmidt-decomposed into the form
\begin{equation}
|\psi\rangle = \sum_a s_a |a\rangle_L |a\rangle_R,
\end{equation}
where $0<s_a<1$ are the Schmidt-weights and $|a\rangle_L$ ($|a\rangle_R$) is the left (right) part 
of the corresponding term in the wave function. If $s_a$ is below a certain threshold, the term 
is discarded. Typically, this threshold is between $10^{-6}$ and $10^{-8}$. For all calculations we 
have checked that the result does not change as the threshold is further decreased. In the 
calculations 
with the lowest thresholds and the largest system sizes, the resulting bond dimensions are on the 
order 
of $10^4$. 

For both, PBC and OBC, our MPS ansatz for the ground state of the finite systems is of the form
\begin{equation}
|\psi\rangle = \sum_{\mu_1\mu_2\dots\mu_L} \ve M^{\mu_1}_1 \cdot \ve M^{\mu_2}_2 \cdots 
\ve M^{\mu_L}_L |\mu_1\rangle |\mu_2\rangle \cdots |\mu_L\rangle, \label{mps_ansatz}
\end{equation}
with $|\mu_j\rangle$  the state of the $j$th site and $\mu_j$ running over the basis states of the 
local Hilbert space at this site. The first and last matrices $\ve M_1^{\mu_1} \in \mathbb C^{1\times 
d}$ 
and $\ve M_L^{\mu_L}\in\mathbb C^{d'\times 1}$ with $d,d'\geq 1$, which means that the periodic 
boundary 
conditions are not built into the MPS. Instead, we explicitly add one long-range term connecting the 
first and the last site of the chain to the matrix product operator which represents the Hamiltonian
with respect to which the ansatz Eq.~(\ref{mps_ansatz}) is optimized. For small systems it has 
been 
checked with exact diagonalization that the ground states of the periodic systems are correct. 
However, the bond dimension needed to reach a similar ground state accuracy is considerably 
higher 
for the periodic boundary conditions.

Once the optimal ground states have been found, which usually is the case after 5-10 sweeps, the 
desired wave function overlaps can be calculated straightforwardly.

For the calculation of the momentum distribution function $n(k)$ we have employed the iDMRG 
algorithm described in Ref. \onlinecite{Kjaell13}. Here we have fixed the bond dimension to 1600. 
Convergence of the ground state requires typically between 50.000 and 100.000 iterations. The 
particle number conservation law has been employed as well. For calculating $n(k)$ we follow 
closely Ref. \onlinecite{Karrasch12}, i.e., we measure the single-particle Green function 
$G_j = \langle c^\dagger_j c_0\rangle$ and compute the Fourier transform $n(k) = \sum_j e^{-i k 
j} G_j$.

{}

\end{document}